\documentclass{article}%
\usepackage{amssymb}
\usepackage{amsfonts}
\usepackage{amsmath}
\usepackage{geometry}
\usepackage{theorem}
\usepackage{graphicx}%
\newtheorem{theorem}{Theorem}[section]

\newtheorem{conjecture}[theorem]{Conjecture}
\newtheorem{corollary}{Corollary}[section]

\newtheorem{definition}[theorem]{Definition}
{\theorembodyfont{\upshape}
\newtheorem{example}[theorem]{Example}
}

\newtheorem{lemma}{Lemma}[section]

{\theorembodyfont{\upshape}
\newtheorem{problem}[theorem]{Problem}
}
\newtheorem{proposition}[theorem]{Proposition}
{\theorembodyfont{\upshape}
\newtheorem{remark}[theorem]{Remark}
}

\newenvironment{proof}[1][Proof]{\noindent\textbf{#1.} }{\ \rule{0.5em}{0.5em}}
\begin{document}

\title{The laplacian of a graph as a density matrix: a basic combinatorial approach
to separability of mixed states}
\author{Samuel L. Braunstein\thanks{ Department of Computer Science, University of
York, \ Heslington, York YO10 5DD, United Kingdom; schmuel@cs.york.ac.uk}
\and Sibasish Ghosh\thanks{Department of Computer Science, University of York;
sibasish@cs.york.ac.uk}
\and Simone Severini\thanks{Department of Mathematics and Department of Computer
Science, University of York; ss54@york.ac.uk}}
\maketitle

\begin{abstract}
We study entanglement properties of mixed density matrices obtained from
combinatorial laplacian matrices of graphs. We observe that some classes of
graphs give arise to entangled (separable) states independently of their labelings.

\smallskip



\end{abstract}
\tableofcontents

\section{Introduction}

Quantum information is a field which has been expanding rapidly due to the
theoretical successes in fast algorithms, super-dense quantum coding, quantum
error correction, teleportation and more. Most of these schemes run off
entanglement in quantum states. Although entanglement in pure state systems is
relatively \textquotedblleft well understood\textquotedblright, this is much
less so in the case of so-called mixed quantum states, which are statistical
mixtures of pure quantum states. In this paper we aim to make some beginning
steps towards improving this situation by focusing our attention to a
restricted class of mixed states. The states we study here may be represented
as graphs in a natural way. We hope that in this manner we may be able to make
powerful statements at least about the class of states represented by what we
call \emph{density matrices of graphs}. We find, for example, that certain
classes of graphs always represent entangled (separable) states. We also find
that a number of considered states have an exactly fractional value of their
concurrence --- a measure of entanglement of formation in small quantum
systems. The representation of a limited class of states by graphs leaves
hints in the expressions we find for possibly natural ways to extend certain
graph-theoretic concepts to more general objects like signed graphs and
weighted graphs.

The paper is divided in six sections. In Section 2, we introduce the notion of
density matrix of a graph. Theorem \ref{pure} characterizes the graphs with
pure density matrices. Theorem \ref{unif} shows that the density matrix of a
graph can be written as a uniform mixture of pure density matrices of graphs.
In Section 3, we consider the von Neumann entropy of density matrices of
graphs. Theorem \ref{minmax} calculates the minimum and maximum von Neumann
entropy that the density matrix of a graph can have, and determines the graphs
for which these values are attained. Theorem \ref{cycles} studies the von
Neumann entropy of the disjoint union of cycles. In Section 4, we discuss
separability. We label the $n=pq$ vertices of a graph by an ordered pair of
indices, where the first index can take $p$ different values and the second
index can take $q$ different values. Theorem \ref{clam} points out that
separability of the density matrix of a graph is generally dependent of the
labelling of the vertices of the graph. This does not hold for complete
graphs, which represent separable states (Lemma \ref{complete1}), and star
graphs, which represent entangled states (Theorem \ref{star}). Theorem
\ref{cecilia} shows that if a graph is a tensor product then its density
matrix is separable, and the converse of it is not necessarily true. After
having introduced the notion of entangled edge, we prove that if all the
entangled edges of a graph on $n=2p$ vertices form a perfect matching, then
the density matrix of the graph is separable in $\mathbb{C}^{2}\otimes
\mathbb{C}^{p}$ (Theorem \ref{mes}). We observe that strongly-regular graphs
and transitive graphs can have entangled or separable density matrix
(Corollary \ref{copetersen}). We calculate the concurrence of all graph on
four vertices representing entangled states. It turns out that for some of
these graphs the value of the concurrence is exactly fractional. In Section 5,
we describe the quantum operations that implement graph transformations like
adding or deleting a vertex or an edge. In Section 6, we state open problems
and conjectures. The paper is relatively self-contained. Our reference on
Graph Theory and Quantum Mechanics are \cite{GR01} and \cite{P93}, respectively.

\section{The density matrix of a graph}

\subsection{Definition}

A \emph{graph} $G=(V,E)$ is a pair defined in the following way: $V$ (or
$V(G)$) is a non-empty and finite set whose elements are called
\emph{vertices}; $E$ (or $E(G)$) is a non-empty set of unordered pairs of
vertices, which are called \emph{edges}. A \emph{loop} is an edge of the form
$\{v_{i},v_{i}\}$, for some vertex $v_{i}$. We assume that $E(G)$ does not
contain only loops. A graph $G$ is said to be \emph{on }$n$ \emph{vertices} if
$\left\vert V(G)\right\vert =n$. The \emph{adjacency matrix} of a graph on $n$
vertices $G$ is an $n\times n$ matrix, denoted by $M(G)$, having rows and
columns labeled by the vertices of $G$, and $ij$-th entry defined as follows:%
\[
\left[  M(G)\right]  _{i,j}=\left\{
\begin{tabular}
[c]{cc}%
$1$ & if $\{v_{i},v_{j}\}\in E(G);$\\
$0$ & if $\{v_{i},v_{j}\}\notin E(G).$%
\end{tabular}
\ \ \right.
\]
Two distinct vertices $v_{i}$ and $v_{j}$ are said to be \emph{adjacent} if
$\{v_{i},v_{j}\}\in E(G)$. The \emph{degree} of a vertex $v_{i}\in V(G)$,
denoted by $d_{G}(v_{i})$, is the number of edges adjacent to $v_{i}$. Two
adjacent vertices are also said to be \emph{neighbours}. The \emph{degree-sum}
of $G$ is defined and denoted by $d_{G}=\sum_{i=1}^{n}d_{G}(v_{i})$. Note that
$d_{G}=2\left\vert E(G)\right\vert $. The \emph{degree} \emph{matrix} of $G$
is an $n\times n$ matrix, denoted by $\Delta(G)$, having $ij$-th entry defined
as follows:%
\[
\left[  \Delta(G)\right]  _{i,j}=\left\{
\begin{tabular}
[c]{cc}%
$d_{G}(v_{i})$ & if $i=j;$\\
0 & if $i\neq j.$%
\end{tabular}
\ \ \right.
\]
The \emph{combinatorial laplacian} \emph{matrix }of a graph $G$ (for short,
\emph{laplacian}) is the matrix $L(G)\overset{def}{=}\Delta(G)-M(G)$. Notice
that $L(G)$ does not change if we add or delete loops from $G$. According to
our definition of graph, $L(G)\neq0$.

\begin{example}
\label{k}Let $I_{n}$ and $J_{n}$ be the $n\times n$ identity matrix and the
$n\times n$ all-ones matrix, respectively. The \emph{complete graph}\ on
$n$\ vertices, denoted by $K_{n}$, is defined to be the graph with adjacency
matrix $\left(  J_{n}-I_{n}\right)  $. Then $L(K_{n})=\left(  n-1\right)
I_{n}-J_{n}+I_{n}=nI_{n}-J_{n}$.
\end{example}

In Standard Quantum Mechanics (that is the Hilbert space formulation of
Quantum Mechanics), the state of a quantum mechanical system associated to the
$n$-dimensional Hilbert space $\mathcal{H}\cong\mathbb{C}^{n}$ is identified
with an $n\times n$ positive semidefinite, trace-one, hermitian matrix, called
a \emph{density matrix}. It is easy to observe that the laplacian of a graph
is symmetric and positive semidefinite. The laplacian of a graph $G$, scaled
by the degree-sum of $G$, has trace one and it is then a density matrix. This
observation leads to the following definition.

\begin{definition}
[Density matrix of a graph]\label{def1}The \emph{density matrix of a graph
}$G$ is the matrix%
\[
\sigma(G)\overset{def}{=}\frac{1}{d_{G}}L(G).
\]

\end{definition}

\begin{example}
The density matrix of $K_{n}$\ (cfr. Example \ref{k}) is $\sigma\left(
K_{n}\right)  =\frac{1}{n\left(  n-1\right)  }\left(  nI_{n}-J_{n}\right)  $.
\end{example}

\subsection{Pure states and mixed states}

Let tr$(A)$ be the trace of a matrix $A$. A density matrix $\rho$ is said to
be \emph{pure} if tr$(\rho^{2})=1$, and \emph{mixed}, otherwise. Theorem
\ref{pure} gives a necessary and sufficient condition on a graph $G$ for
$\sigma(G)$ to be pure. We first provide some terminology and state an easy
lemma. A graph $G$ is said to have $k$ \emph{components}, $G_{1}%
,G_{2},...,G_{k}$, and in such a case we write $G=G_{1}\uplus G_{2}%
\uplus\cdots\uplus G_{k}$, if there is an ordering of $V(G)$, such that $M(G)=%
{\textstyle\bigoplus_{i=1}^{k}}
M(G_{i})$. When $k=1$, $G$ is said to be \emph{connected}. From now on, we
denote by $\lambda_{1}(A),\lambda_{2}(A),...,\lambda_{k}(A)$ the $k$ different
eigenvalues of an Hermitian matrix $A$ in increasing order. The set of the
eigenvalues of $A$ together with their multiplicities is called
\emph{spectrum} of $A$.

\begin{lemma}
\label{eig}The density matrix of a graph $G$ has a zero eigenvalue whose
multiplicity is equal to the number of components of $G$.
\end{lemma}

\begin{proof}
Given a graph $G$, it is a direct consequence of Definition \ref{def1} that
$\lambda_{i}(\sigma(G))=\frac{\lambda_{i}(L(G))}{d_{G}}$. It is well-known
that $L(G)$ has a zero eigenvalue whose multiplicity is equal to the number of
components of $G$ \cite{GR01}.
\end{proof}

\begin{theorem}
\label{pure}The density matrix of a graph $G$ is pure if and only if $G=K_{2}
$ or $G=K_{2}\uplus v_{1}\uplus v_{2}\uplus\cdots\uplus v_{l}$, for some
vertices $v_{1},v_{2},...,v_{l}$. (These vertices are with or without loops.)
\end{theorem}

\begin{proof}
Let $G$ be a graph on $n$ vertices. Suppose that $\sigma(G)$ is pure. By the
definition of pure density matrix, the different eigenvalues of $\sigma(G)$
are $\lambda_{1}\left(  \sigma(G)\right)  =0$ and $\lambda_{2}\left(
\sigma(G)\right)  =1$. Moreover, $\lambda_{1}\left(  \sigma(G)\right)  =0$ has
multiplicity $\left(  n-1\right)  $. Then, by Lemma \ref{eig}, the number of
components of $G$ is $\left(  n-1\right)  $. Since $\left\vert V(G)\right\vert
=n$, it follows that $G=K_{2}\uplus v_{1}\uplus v_{2}\uplus\cdots\uplus v_{l}%
$, where $l=n-2$.
\end{proof}

\bigskip

The next definition is based on the theorem.

\begin{definition}
[Pure density matrix of a graph]Let $G$ be a graph on $n\geq2$ vertices. The
density matrix of graph $G$ is said to be \emph{pure} if $G=K_{2}\uplus
v_{1}\uplus v_{2}\uplus\cdots\uplus v_{n-2}$, for some vertices $v_{1}%
,v_{2},...,v_{n-2}$. (These vertices are with or without loops.)
\end{definition}

\begin{example}
The density matrix
\[
\sigma(K_{2})=\left[
\begin{array}
[c]{rr}%
\frac{1}{2} & -\frac{1}{2}\\
-\frac{1}{2} & \frac{1}{2}%
\end{array}
\right]
\]
is pure. In fact, $\lambda_{1}\left(  \sigma(K_{2})\right)  =0$\ and
$\lambda_{2}\left(  \sigma(K_{2})\right)  =1$.
\end{example}

A graph $H$ is said to be a \emph{factor} of a graph $G$, if $V(H)=V(G)$ and
there exists a graph $H^{\prime}$ such that $V(H^{\prime})=V(G)$ and
$M(G)=M(H)+M(H^{\prime})$.

\begin{theorem}
\label{unif}The density matrix of a graph is a uniform mixture of pure density matrices.
\end{theorem}

\begin{proof}
Let $G$ be a graph on $n$ vertices $v_{1},v_{2},...,v_{n}$, having edges
$\{v_{i_{1}},v_{j_{1}}\},\{v_{i_{2}},v_{j_{2}}\},...,\{v_{i_{m}},v_{j_{m}}\}$,
where $1\leq i_{1},j_{1},i_{2},j_{2},...,i_{m},j_{m}\leq n$. Let
$H_{i_{k}j_{k}}$ be the factor of $G$ such that
\begin{equation}%
\begin{tabular}
[c]{c}%
$\left[  M(H_{i_{k}j_{k}})\right]  _{u,w}=\left\{
\begin{tabular}
[c]{ll}%
$1$ & if $u=i_{k}$ and $w=j_{k}$ or $w=i_{k}$ and $u=j_{k}$;\\
$0$ & otherwise.
\end{tabular}
\ \ \right.  $%
\end{tabular}
\ \ \label{factor}%
\end{equation}
By Theorem \ref{pure}, the density matrix $\sigma(H_{i_{k}j_{k}})=\frac{1}%
{2}\left(  \Delta(H_{i_{k}j_{k}})-M(H_{i_{k}j_{k}})\right)  $ is pure. Since
\[%
\begin{tabular}
[c]{lll}%
$\Delta(G)=%
{\displaystyle\sum\limits_{k=1}^{m}}
\Delta(H_{i_{k}j_{k}})$ & and & $M(G)=%
{\displaystyle\sum\limits_{k=1}^{m}}
M(H_{i_{k}j_{k}}),$%
\end{tabular}
\ \
\]
we can write
\begin{equation}
\sigma\left(  G\right)  =\frac{1}{2m}\left(  \Delta(G)-M(G)\right)  =\frac
{1}{m}%
{\displaystyle\sum\limits_{k=1}^{m}}
\sigma(H_{i_{k}j_{k}}), \label{factor1}%
\end{equation}
which is then a uniform mixture of pure density matrices.
\end{proof}

\begin{example}
Consider a graph $G$\ defined as follows: $V(G)=\{1,2,3\}$\ and
$E(G)=\{e=\{1,2\},f=\{2,3\}\}$. Then%
\[%
\begin{tabular}
[c]{lll}%
$M(H_{1_{e}2_{e}})=\left[
\begin{array}
[c]{ccc}%
0 & 1 & 0\\
1 & 0 & 0\\
0 & 0 & 0
\end{array}
\right]  ,$ &  & $M(H_{2_{f}3_{f}})=\left[
\begin{array}
[c]{ccc}%
0 & 0 & 0\\
0 & 0 & 1\\
0 & 1 & 0
\end{array}
\right]  $%
\end{tabular}
\]
and
\[
\sigma(G)=\frac{1}{2}\left(  \sigma(H_{1_{e}2_{e}})+\sigma(H_{2_{f}3_{f}%
})\right)  =\left[
\begin{array}
[c]{rrr}%
\frac{1}{4} & -\frac{1}{4} & 0\\
-\frac{1}{4} & \frac{1}{2} & -\frac{1}{4}\\
0 & -\frac{1}{4} & \frac{1}{4}%
\end{array}
\right]  .
\]

\end{example}

\section{Von Neumann entropy}

The \emph{von Neumann entropy} of an $n\times n$ density matrix $\rho$ is
$S(\rho)\overset{def}{=}-\sum_{i=1}^{n}\lambda_{i}\left(  \rho\right)
\log_{2}\lambda_{i}\left(  \rho\right)  $. It is conventional to define
$0\log_{2}0=0$. The von Neumann entropy is a quantitative measure of mixedness
of the density matrix $\rho$.

\begin{remark}
The $q$\emph{-entropy} of an $n\times n$ density matrix $\rho$ is $($%
tr$(\rho^{q}))^{1/q}$. $q$-entropies are a family of measures of mixedness for
density matrices. In general, in the limit $q\rightarrow\infty$, $q$-entropies
depend only on the largest eigenvalues of $\rho$, and we have $\lim
_{q\rightarrow\infty}($tr$(\rho^{q})^{1/q}=\lambda_{n}(\rho)$. This eigenvalue
can be considered itself as a measure of mixedness \cite{BCPP02}. If $\rho$ is
the density matrix of a graph a tight upper-bound on $\lambda_{n}(\rho)$ is
known \cite{SHK02}.
\end{remark}

\subsection{Maximum and minimum}

\begin{theorem}
\label{minmax}Let $G$ be a graph on $n$ vertices. Then

\begin{enumerate}
\item $\max\limits_{G}S(\sigma(G))=\log_{2}\left(  n-1\right)  =S(\sigma
(K_{n}))$;

\item $\min\limits_{G}S(\sigma(G))=0$ and this value is attained if
$\sigma(G)$ is pure.
\end{enumerate}
\end{theorem}

\begin{proof}
(Proof of 1) By Lemma \ref{eig}, $\sigma(G)$ has an eigenvalue zero with
multiplicity at least one. Since $G$ is on $n$ vertices, the support of
$\sigma(G)$ has dimension less or equal to $\left(  n-1\right)  $. Any
$n\times n$ density matrix having dimension of support less or equal to
$\left(  n-1\right)  $, can not have von Neumann entropy greater than
$\log_{2}\left(  n-1\right)  $. The eigenvalues of $\sigma(K_{n})$ are
$\lambda_{1}\left(  \sigma(K_{n})\right)  =0$, with multiplicity $1$, and
$\lambda_{2}\left(  \sigma(K_{n})\right)  =\tfrac{1}{\left(  n-1\right)  }$,
with multiplicity $\left(  n-1\right)  $. Then
\[
S\left(  \sigma(K_{n})\right)  =-%
{\displaystyle\sum}
\frac{1}{n-1}\log_{2}\frac{1}{n-1}=\log_{2}\left(  n-1\right)  .
\]
(Proof of 2) Since $G$ is a graph on $n$ vertices, the maximum multiplicity of
the zero eigenvalue of $\sigma(G)$ is $\left(  n-1\right)  $; the other
eigenvalue of $\sigma(G)$ is necessarily one. This is the case when
$\sigma(G)$ is pure. When $\sigma(G)$ is pure, $S\left(  \sigma(G)\right)  =0$.
\end{proof}

\subsection{Regular graphs}

Two graphs $G$ and $H$ are said to be $L$-\emph{cospectral} if $L(G)$ and
$L(H)$ have the same spectrum; $\sigma$-\emph{cospectral} if $\sigma(G)$ and
$\sigma(H)$ have the same spectrum. Two graphs $G$ and $H$ are said to be
\emph{isomorphic}, and in such a case we write $G\cong H$, if there is an
\emph{isomorphism }between $V(G)$ and $V(H)$, that is there is a permutation
matrix $P$, such that $PM(G)P^{\intercal}=M(H)$. If $G\cong H$ then $G$ and
$H$ are $L$-cospectral and $\sigma$-cospectral, but the converse is not
necessarily true. Two graphs are $L$-cospectral and $\sigma$-cospectral if and
only if they have the same degree sum. Now, a graph is said to be
\emph{regular} if each of its vertices has the same degree. A $d$%
\emph{-regular} graph is a regular graph whose degree of the vertices is $d$.
If $G$ is $d$-regular graph on $n$ vertices, then $\lambda_{i}(L(G))=d-\lambda
_{i}(M(G))$ and $\lambda_{i}(\sigma(G))=\tfrac{d-\lambda_{i}(M(G))}{dn}$,
because $d_{G}=dn$. So, $G$ and $H$ are $L$-cospectral $d$-regular graphs if
and only if they are $\sigma$-cospectral. Now, let us consider a $d$-regular
graph $G$. Let us write $\sigma_{i}=\lambda_{i}(\sigma(G))$ and $\mu
_{i}=\lambda_{i}(M(G))$. Let $m_{i}$ be the multiplicity of the $i$-th
eigenvalue of $M(G)$. This is also the multiplicity of the $i$-th eigenvalue
of $\sigma(G)$, given that $G$ is regular. The von Neumann entropy of $G$ is
then given by%

\begin{align*}
S(\sigma(G))  &  =-%
{\displaystyle\sum\limits_{i=1}^{k}}
m_{i}\left(  \sigma_{i}\log_{2}\sigma_{i}\right)  =-\frac{1}{d\cdot n}%
{\displaystyle\sum\limits_{i=1}^{k}}
m_{i}\left[  \left(  d-\mu_{i}\right)  \log_{2}\left(  d-\mu_{i}\right)
\right]  +\frac{\log_{2}\left(  d\cdot n\right)  }{d\cdot n}%
{\displaystyle\sum\limits_{i=1}^{k}}
m_{i}\left(  d-\mu_{i}\right) \\
&  =-\frac{1}{d\cdot n}%
{\displaystyle\sum\limits_{i=1}^{k}}
m_{i}\left[  \left(  d-\mu_{i}\right)  \log_{2}\left(  d-\mu_{i}\right)
\right]  +\log_{2}(d\cdot n),
\end{align*}

\subsection{Cycles}

Let $\Gamma$ be a finite group. Let $S\subset\Gamma$ be a subset of $\Gamma$,
such that: the set $S$ does not contain the identity element; an element $s\in
S$ if and only if $s^{-1}\in S$. Let $\rho_{reg}(g)$ be the (left) regular
permutation representation of an element $g\in\Gamma$. The (\emph{left})
\emph{Cayley graph} of $\Gamma$ \emph{with respect to} $S$, denoted by
$X(\Gamma,S)$, is defined to be the graph with adjacency matrix $M(X(\Gamma
,S))=%
{\textstyle\sum\nolimits_{s\in S}}
\rho_{reg}(s)$. Notice that $X(\Gamma,S)$ is connected if and only if $S$
generates $\Gamma$.

\begin{example}
Let $\Gamma=\mathbb{Z}_{n}$ be the group of the integers modulo $n$\ and let
$S=\{1,n-1\}\subset\Gamma$. Let $G\cong X(\Gamma,S)$. Then $M(G)=\rho
_{reg}(1)+\rho_{reg}(n-1)$. Since $S$\ generates $\Gamma$, the graph $G$\ is
connected. The $n$-\emph{cycle}, denoted by $C_{n}$, is a graph on
$n$\ vertices $v_{1},v_{2},...,v_{n}$\ and with $n$\ edges $\{v_{1}%
,v_{2}\},\{v_{2},v_{3}\},...,\{v_{n-1},v_{n}\},\{v_{n},v_{1}\}$. Hence,
$G\cong C_{n}$.
\end{example}

\begin{theorem}
\label{cycles}Let $G_{k}=X(\Gamma,S_{k})$ be a Cayley graph, where
$\Gamma=\mathbb{Z}_{n}$ and $S_{k}=\{k,n-k\}\subset\Gamma$. Then

\begin{enumerate}
\item $\max_{G_{k}}S(\sigma(G_{k}))=S(\sigma(C_{n}))$, that is when
$\gcd(k,n)=1$;

\item $\min_{G_{k}}S(\sigma(G_{k}))=S(\sigma(C_{\frac{n}{2}}))$.
\end{enumerate}
\end{theorem}

\begin{proof}
(Proof of 1) We begin by observing that, given $S_{k}=\{k,n-k\}$, with
$k=n/p$, $G_{k}=C_{p}\uplus C_{p}\uplus\cdots\uplus C_{p}$, where $C_{p}$ is
repeated $k$-times. This indicates that the eigenvalues of $M\left(
G_{k}\right)  $ are the eigenvalues of $M\left(  C_{p}\right)  $, each
repeated $k$ times: $\lambda_{j}\left(  \sigma(G_{k})\right)  =\tfrac
{2p\cdot\lambda_{j}\left(  \sigma(C_{p})\right)  }{2n}$, where $1\leq j\leq
p$, and each $\lambda_{j}\left(  \sigma(G_{k})\right)  $ has multiplicity $k$.
Since, it is well-known that $\lambda_{j}\left(  M(C_{p})\right)
=2\cos\left(  2\pi j/p\right)  $, where $1\leq j\leq p$, we have $\lambda
_{j}\left(  \sigma(C_{p})\right)  =\tfrac{2-2\cos\left(  2\pi j/p\right)
}{2p}=\tfrac{2\sin^{2}\left(  \pi j/p\right)  }{p}$ and $\lambda_{j}\left(
\sigma(G_{k})\right)  =\tfrac{2\sin^{2}\left(  \pi j/p\right)  }{n}$. By
writing
\[
A_{p}\left(  j\right)  =\sin^{2}\left(  \frac{\pi j}{p}\right)  \log
_{2}\left(  \sin^{2}\left(  \frac{\pi j}{p}\right)  \right)  ,
\]
the von Neumann entropy of $\sigma(G_{k})$ is given by%
\[
S(\sigma(G_{k}))=-k%
{\displaystyle\sum\limits_{j=1}^{p}}
\frac{2}{n}A_{p}\left(  j\right)  =\log_{2}n-1-\frac{2}{n}%
{\displaystyle\sum\limits_{j=1}^{p}}
A_{p}\left(  j\right)  .
\]
Because we do not have any closed form of the series $-%
{\displaystyle\sum\limits_{j=1}^{p}}
A_{p}\left(  j\right)  $, we use the following approximation, which is very
good for large $p$:
\begin{equation}%
\begin{tabular}
[c]{lll}%
$-%
{\displaystyle\sum\limits_{j=1}^{p}}
A_{p}\left(  j\right)  \simeq-\frac{p}{\pi}%
{\displaystyle\int\limits_{0}^{\pi}}
\sin^{2}x\log_{2}\left(  \sin^{2}x\right)  dx=p\cdot C,$ & where & $C=\left(
1-\frac{\log_{2}e}{2}\right)  \simeq0.2787.$%
\end{tabular}
\ \label{app}%
\end{equation}
If $p=1,2$, $-%
{\displaystyle\sum\limits_{j=1}^{p}}
A_{p}\left(  j\right)  =0$. So, if $p=1,2$ and $n$ is even, $S(\sigma
(G_{n/2}))=\log_{2}n-1$. With the use of Equation \ref{app}, we obtain
$S(\sigma(G_{k}))\simeq\log_{2}n-1+2C/k$. If $l=n/q$ then $S(\sigma
(G_{k}))-S(\sigma(G_{l}))\simeq2C\left(  \frac{1}{k}-\frac{1}{l}\right)  $. It
follows that: $S(\sigma(G_{k}))>S(\sigma(G_{l}))$ if $l>k$; $S(\sigma
(G_{k}))<S(\sigma(G_{l}))$ if $l<k$. When $k=1$ then $G_{k}=C_{n}$. Therefore,
$S(\sigma(C_{n}))>S(\sigma(G_{l}))$, for all $l>1$. (Proof of 2) By the
reasoning above, it is sufficient to observe that $S(\sigma(G_{\frac{n}{2}%
}))=\log_{2}n-1$.
\end{proof}

\begin{example}
In the table below, the values of the von Neumann entropy of the Cayley graphs
$X(\mathbb{Z}_{12},S)$, where $\left\vert S\right\vert =2$ are given:%
\[%
\begin{tabular}
[c]{|l|l|}\hline
$\ \ \ \ \ \ \ \ \ G$ & $S(\sigma(G))$\\\hline
$X(\mathbb{Z}_{12},\left\{  1,11\right\}  )$ & $3.571$\\\hline
$X(\mathbb{Z}_{12},\left\{  2,10\right\}  )$ & $3.126$\\\hline
$X(\mathbb{Z}_{12},\left\{  3,9\right\}  )$ & $3.084$\\\hline
$X(\mathbb{Z}_{12},\left\{  4,8\right\}  )$ & $3.000$\\\hline
$X(\mathbb{Z}_{12},\left\{  6\right\}  )$ & $2.585$\\\hline
\end{tabular}
\
\]

\end{example}

\section{Separability}

Let $\emph{S}_{A}$ and $\emph{S}_{B}$ be two quantum mechanical systems,
associated to the $p$-dimensional and $q$-dimensional Hilbert spaces
$\mathcal{H}_{A}\cong\mathbb{C}_{A}^{p}$ and $\mathcal{H}_{B}\cong
\mathbb{C}_{B}^{q}$, respectively. The composite system $\emph{S}_{AB}$, which
consists of the subsystems $\emph{S}_{A}$ and $\emph{S}_{B}$, is associated to
the Hilbert space $\mathbb{C}_{A}^{p}\otimes\mathbb{C}_{B}^{q}$, where
\textquotedblleft$\otimes$\textquotedblright\ denotes tensor product. The
density matrix $\rho_{AB}$ of $\emph{S}_{AB}$ is said to be \emph{separable}
if $\rho_{AB}=%
{\textstyle\sum\nolimits_{i=1}^{n}}
\omega_{i}\rho_{A}^{\left(  i\right)  }\otimes\rho_{B}^{\left(  i\right)  }$,
where $\omega_{i}\geq0$, for every $i=1,2,...,n$, and $%
{\textstyle\sum\nolimits_{i=1}^{n}}
\omega_{i}=1$; $\rho_{A}^{\left(  i\right)  }$ and $\rho_{B}^{\left(
i\right)  }$ are density matrices acting on $\mathcal{H}_{A}$ and
$\mathcal{H}_{B}$, respectively. A density matrix $\rho_{AB}$ is said to be
\emph{entangled} if it is not separable. In Dirac notation, a unit vector in a
Hilbert space $\mathcal{H}\cong\mathbb{C}^{n}$ is denoted by $\left\vert
\psi\right\rangle $, where $\psi$ is a label; given the vectors $\left\vert
\varphi\right\rangle ,\left\vert \psi\right\rangle \in\mathcal{H}$, the linear
functional sending $\left\vert \psi\right\rangle $ to the inner product
$\left\langle \varphi|\psi\right\rangle $ is denoted by $\left\langle
\varphi\right\vert $. We write $\left\vert \psi\right\rangle \left\vert
\varphi\right\rangle $ for the tensor product $\left\vert \psi\right\rangle
\otimes\left\vert \varphi\right\rangle $. A vector of the form $\left\vert
\psi\right\rangle \left\vert \varphi\right\rangle $ is called \emph{product
state}. For any unit vector $\left\vert \psi\right\rangle \in\mathcal{H}$, the
projector on $\left\vert \psi\right\rangle $ is the hermitian matrix
$\left\vert \psi\right\rangle \left\langle \psi\right\vert $ which we denote
by $P\left[  \left\vert \psi\right\rangle \right]  $.

\subsection{Tensor product of graphs}

The \emph{tensor product of graphs} $G$ and $H$ (also known in literature as
strong product, cardinal product, \emph{etc.}), denoted by $G\otimes H$, is
the graph whose adjacency matrix is $M(G\otimes H)=M(G)\otimes M(H)$
\cite{IK00} . Whenever we consider a graph $G\otimes H$, where $G$ is on $p$
vertices and $H$ is on $q$ vertices, the separability of $\sigma(G\otimes H)$
is described with respect to the Hilbert space $\mathcal{H}_{G}\otimes
\mathcal{H}_{H}$, where $\mathcal{H}_{G}$ is the space spanned by the
orthonormal basis $\left\{  \left\vert u_{1}\right\rangle ,\left\vert
u_{2}\right\rangle ,...,\left\vert u_{p}\right\rangle \right\}  $ associated
to $V(G)$, and $\mathcal{H}_{H}$ is the space spanned by the orthonormal basis
$\left\{  \left\vert w_{1}\right\rangle ,\left\vert w_{2}\right\rangle
,...,\left\vert w_{q}\right\rangle \right\}  $ associated to $V(H)$. The
vertices of $G\otimes H$ are taken as $u_{1}w_{1},u_{1}w_{2},...,u_{1}%
w_{q},u_{2}w_{1},u_{2}w_{2},...,u_{p}w_{q}$. We associate $\left\vert
u_{1}\right\rangle \left\vert w_{1}\right\rangle $ to $u_{1}w_{1}$,
$\left\vert u_{1}\right\rangle \left\vert w_{2}\right\rangle $ to $u_{1}w_{2}%
$,..., $|u_{p}\left\vert w_{q}\right\rangle $ to $u_{p}w_{q}$ . In conjunction
with this, whenever we talk about separability of any graph $G$ on $n$
vertices, $v_{1},v_{2},...,v_{n}$, we consider it in the space $\mathbb{C}%
^{p}\otimes\mathbb{C}^{q}$, where $n=pq$. The vectors $\left\vert
v_{1}\right\rangle ,\left\vert v_{2}\right\rangle ,...,\left\vert
v_{n}\right\rangle $ are taken as follows: $\left\vert v_{1}\right\rangle
=\left\vert u_{1}\right\rangle \left\vert w_{1}\right\rangle ,\left\vert
v_{2}\right\rangle =\left\vert u_{1}\right\rangle \left\vert w_{2}%
\right\rangle ,...,\left\vert v_{q}\right\rangle =\left\vert u_{1}%
\right\rangle \left\vert w_{q}\right\rangle $, $\left\vert v_{q+1}%
\right\rangle =\left\vert u_{2}\right\rangle \left\vert w_{1}\right\rangle
,\left\vert v_{q+2}\right\rangle =\left\vert u_{2}\right\rangle \left\vert
w_{2}\right\rangle ,...,\left\vert v_{2q}\right\rangle =\left\vert
u_{2}\right\rangle \left\vert w_{q}\right\rangle ,...,|v_{pq}\rangle
=|\left\vert u_{p}\right\rangle \left\vert w_{q}\right\rangle $. We make use
of the notion of \emph{partial transpose} of a density matrix. Let us consider
a $pq\times pq$ density matrix $\rho_{AB}$ acting on $\mathbb{C}_{A}%
^{p}\otimes\mathbb{C}_{B}^{q}$. Let $\left\{  \left\vert v_{1}\right\rangle
,\left\vert v_{2}\right\rangle ,...,\left\vert v_{p}\right\rangle \right\}  $
and $\left\{  \left\vert w_{1}\right\rangle ,\left\vert w_{2}\right\rangle
,...,\left\vert w_{q}\right\rangle \right\}  $ be orthonormal bases of
$\mathbb{C}_{A}^{p}$ and $\mathbb{C}_{B}^{q}$, respectively. The \emph{partial
transpose} of $\rho_{AB}$ with respect to the system $\emph{S}_{B}$ is the
$pq\times pq$ matrix, denoted by $\rho_{AB}^{\intercal_{B}}$, and with
$\left(  i,j;i^{\prime},j^{\prime}\right)  $-th entry defined as follows:
$[\rho_{AB}^{\intercal_{B}}]_{i,j;i^{\prime},j^{\prime}}=\langle v_{i}|\langle
w_{j^{\prime}}|\rho_{AB}|v_{i^{\prime}}\rangle|w_{j}\rangle$, where $1\leq
i,i^{\prime}\leq p$ and $1\leq j,j^{\prime}\leq q$. Regarding separability of
$\rho_{AB}$ we have the following criterion \cite{P96, HHH96}:

\noindent\textbf{(Peres-Horodecki Criterion (PH)) }\label{pht}If $\rho$ is a
density matrix acting on $\mathbb{C}^{2}\otimes\mathbb{C}^{2}$ or
$\mathbb{C}^{2}\otimes\mathbb{C}^{3}$, then $\rho$ is separable if and only if
$\rho^{\intercal_{B}}$ is positive semidefinite.

\begin{theorem}
\label{clam}Let $G$ and $H$ be two graphs on $n=p\cdot q$ vertices. If
$\sigma(G)$ is entangled in $\mathbb{C}^{p}\otimes\mathbb{C}^{q}$ and $G\cong
H$ then $\sigma(H)$ is not necessarily entangled in $\mathbb{C}^{p}%
\otimes\mathbb{C}^{q}$.
\end{theorem}

\begin{proof}
Let $G$ be a graph on the vertices $1,2,3$ and $4$, having edges
$\{1,2\},\{2,3\}$ and $\{3,4\}$. We associate to $G$ the following orthonormal
basis: $\{\left\vert 1\right\rangle =\left\vert 1\right\rangle _{A}\left\vert
1\right\rangle _{B},\left\vert 2\right\rangle =\left\vert 2\right\rangle
_{A}\left\vert 1\right\rangle _{B},\left\vert 3\right\rangle =\left\vert
1\right\rangle _{A}\left\vert 2\right\rangle _{B},\left\vert 4\right\rangle
=\left\vert 2\right\rangle _{A}\left\vert 2\right\rangle _{B}\}$. In terms of
this basis$\,$%
\[
\left(  \sigma(G)\right)  ^{\intercal_{B}}=\frac{1}{6}\left[
\begin{array}
[c]{rrrr}%
1 & -1 & 0 & -1\\
-1 & 2 & 0 & 0\\
0 & 0 & 2 & -1\\
-1 & 0 & -1 & 1
\end{array}
\right]  ,
\]
with spectrum $\{[1/2],[1/6],[(1+\sqrt{2})/6],[(1-\sqrt{2})/6]\}$. Since the
last eigenvalue is negative, by the PH criterion, $\sigma(P_{4})$ is
entangled. Consider the graph $H\cong G$. The edges of $H$ are\thinspace
$\{1,4\},\{4,3\}$ and $\{3,2\}$. We associate to $H$ the above orthonormal
basis. We then have%
\[
\left(  \sigma(H)\right)  ^{\intercal_{B}}=\frac{1}{6}\left[
\begin{array}
[c]{rrrr}%
1 & 0 & 0 & -1\\
0 & 1 & -1 & 0\\
0 & -1 & 2 & -1\\
-1 & 0 & -1 & 2
\end{array}
\right]  =\sigma(H),
\]
and so $\sigma(H)$ is separable.
\end{proof}

\begin{lemma}
\label{prod}The density matrix of the tensor product of two graphs is separable.
\end{lemma}

\begin{proof}
Let $G$ be a graph on $n$ vertices, $v_{1},v_{2},...,v_{n}$, and $m$ edges,
$\{v_{i_{1}},v_{j_{1}}\},\{v_{i_{2}},v_{j_{2}}\},...,\{v_{i_{m}},v_{j_{m}}\}$,
where $1\leq i_{1},j_{1},i_{2},j_{2},...,i_{m},j_{m}\leq n$. Let $G^{\prime}$
be a graph on $p$ vertices, $v_{1},v_{2},...,v_{p}$, and $q$ edges,
$\{v_{s_{1}}^{\prime},v_{t_{1}}^{\prime}\},\{v_{s_{2}}^{\prime},v_{t_{2}%
}^{\prime}\},...,\{v_{s_{q}}^{\prime},v_{t_{q}}^{\prime}\}$, where $1\leq
s_{1},t_{1},s_{2},t_{2},...,s_{q},t_{q}\leq p$. By Theorem \ref{unif}
(Equation \ref{factor1}), we can write
\[
\sigma\left(  G\right)  =\frac{1}{m}%
{\displaystyle\sum\limits_{k=1}^{m}}
\sigma(H_{i_{k}j_{k}})\text{ and }\sigma\left(  G^{\prime}\right)  =\frac
{1}{q}%
{\displaystyle\sum\limits_{l=1}^{q}}
\sigma(L_{s_{l}t_{l}}),
\]
where $H_{i_{k}j_{k}}$ and $L_{s_{l}t_{l}}$ are defined according to Equation
\ref{factor}. So,%
\begin{align}
\sigma\left(  G\otimes G^{\prime}\right)   &  =\frac{1}{d_{G\otimes G^{\prime
}}}\left[  \Delta\left(  G\otimes G^{\prime}\right)  -M(G\otimes G^{\prime
})\right] \nonumber\\
&  =\frac{1}{d_{G\otimes G^{\prime}}}%
{\displaystyle\sum\limits_{k=1}^{m}}
{\displaystyle\sum\limits_{l=1}^{q}}
\left[  \Delta\left(  H_{i_{k}j_{k}}\otimes L_{s_{l}t_{l}}\right)
-M(H_{i_{k}j_{k}}\otimes L_{s_{l}t_{l}})\right] \\
&  =\frac{1}{m\cdot q}%
{\displaystyle\sum\limits_{k=1}^{m}}
{\displaystyle\sum\limits_{l=1}^{q}}
\sigma\left(  H_{i_{k}j_{k}}\otimes L_{s_{l}t_{l}}\right) \nonumber\\
&  =\frac{1}{m\cdot q}%
{\displaystyle\sum\limits_{k=1}^{m}}
{\displaystyle\sum\limits_{l=1}^{q}}
\frac{1}{2}[\sigma^{+}\left(  H_{i_{k}j_{k}}\right)  \otimes\sigma\left(
L_{s_{l}t_{l}}\right)  +\sigma\left(  H_{i_{k}j_{k}}\right)  \otimes\sigma
^{+}\left(  L_{s_{l}t_{l}}\right)  ],
\end{align}
where%
\[%
\begin{tabular}
[c]{lll}%
$\sigma^{+}\left(  H_{i_{k}j_{k}}\right)  \overset{def}{=}\Delta\left(
H_{i_{k}j_{k}}\right)  -\sigma\left(  H_{i_{k}j_{k}}\right)  $ & and &
$\sigma^{+}\left(  L_{s_{l}t_{l}}\right)  \overset{def}{=}\Delta\left(
L_{s_{l}t_{l}}\right)  -\sigma\left(  L_{s_{l}t_{l}}\right)  .$%
\end{tabular}
\]
Notice that $\sigma^{+}\left(  H_{i_{k}j_{k}}\right)  $ and $\sigma^{+}\left(
L_{s_{l}t_{l}}\right)  $ are density matrices. Let
\[%
\begin{tabular}
[c]{lll}%
$\sigma^{+}\left(  G\right)  \overset{def}{=}\frac{1}{m}%
{\displaystyle\sum\limits_{k=1}^{m}}
\sigma^{+}(H_{i_{k}j_{k}})$ & and & $\sigma^{+}\left(  G^{\prime}\right)
\overset{def}{=}\frac{1}{q}%
{\displaystyle\sum\limits_{l=1}^{q}}
\sigma^{+}(L_{s_{l}t_{l}}).$%
\end{tabular}
\]
Then%
\begin{equation}
\sigma\left(  G\otimes G^{\prime}\right)  =\frac{1}{2}[\sigma\left(  G\right)
\otimes\sigma^{+}(G^{^{\prime}})+\sigma^{+}\left(  G\right)  \otimes
\sigma(G^{^{\prime}})]. \label{ggprime2}%
\end{equation}
Since each of $\sigma\left(  G\right)  ,\sigma^{+}(G^{^{\prime}}),\sigma
^{+}\left(  G\right)  $ and $\sigma(G^{^{\prime}})$ is a uniform mixture of
density matrices, then $\sigma\left(  G\otimes G^{\prime}\right)  $ is separable.
\end{proof}

\bigskip

We associate to the vertices, $v_{1},v_{2},...,v_{n}$, of a graph $G$ an
orthonormal basis $\left\{  \left\vert v_{1}\right\rangle ,\left\vert
v_{2}\right\rangle ,...,\left\vert v_{n}\right\rangle \right\}  $. In terms of
this basis, the $uw$-th elements of the matrices $\sigma\left(  H_{i_{k}j_{k}%
}\right)  $ and $\sigma^{+}\left(  H_{i_{k}j_{k}}\right)  $ are given by
$\left\langle v_{u}|\sigma\left(  H_{i_{k}j_{k}}\right)  |v_{w}\right\rangle $
and $\left\langle v_{u}|\sigma^{+}\left(  H_{i_{k}j_{k}}\right)
|v_{w}\right\rangle $, respectively. In this basis
\[%
\begin{tabular}
[c]{lll}%
$\sigma\left(  H_{i_{k}j_{k}}\right)  =P[\frac{1}{\sqrt{2}}\left(  \left\vert
v_{i_{k}}\right\rangle -\left\vert v_{j_{k}}\right\rangle \right)  ]$ & and &
$\sigma^{+}\left(  H_{i_{k}j_{k}}\right)  =P[\frac{1}{\sqrt{2}}\left(
\left\vert v_{i_{k}}\right\rangle +\left\vert v_{j_{k}}\right\rangle \right)
].$%
\end{tabular}
\]

\begin{lemma}
\label{complete}For any $n=p\cdot q$, the density matrix $\sigma(K_{n})$ is
separable in $\mathbb{C}^{p}\otimes\mathbb{C}^{q}$.
\end{lemma}

\begin{proof}
Let $v_{1},v_{2},...,v_{n}$ be the vertices of $K_{n}$, with $n=p\cdot q$. Let
us consider the following two orthonormal bases $\left\{  \left\vert
u_{1}\right\rangle ,\left\vert u_{2}\right\rangle ,...,\left\vert
u_{p}\right\rangle \right\}  $ and $\left\{  \left\vert w_{1}\right\rangle
,\left\vert w_{2}\right\rangle ,...,\left\vert w_{q}\right\rangle \right\}  $
of $\mathbb{C}^{p}$ and $\mathbb{C}^{q}$, respectively. For all $i=1,2,...,n$,
we then write $\left\vert v_{i}\right\rangle =\left\vert u_{s+1}\right\rangle
\left\vert w_{s^{\prime}}\right\rangle $, where $i=sq+s^{\prime}$, $0\leq
s\leq p-1$ and $1\leq s^{\prime}\leq q$. By making use of this basis, we can
write $\sigma\left(  H_{i_{k}j_{k}}\right)  =P[\frac{1}{\sqrt{2}}(\left\vert
u_{s_{k}+1}\right\rangle |w_{s_{k}^{\prime}}\rangle-\left\vert u_{t_{k}%
+1}\right\rangle |w_{t_{k}^{\prime}}\rangle)]$, where $i_{k}=s_{k}%
q+s_{k}^{\prime}$, $j_{k}=t_{k}q+t_{k}^{\prime}$, $0\leq s_{k},t_{k}\leq p-1$
and $0\leq s_{k}^{\prime},t_{k}^{\prime}\leq q$. By Equation \ref{factor1},
\[
\sigma\left(  K_{n}\right)  =\frac{1}{m}%
{\displaystyle\sum\limits_{k=1}^{m}}
P[\frac{1}{\sqrt{2}}(\left\vert u_{s_{k}+1}\right\rangle |w_{s_{k}^{\prime}%
}\rangle-\left\vert u_{t_{k}+1}\right\rangle |w_{t_{k}^{\prime}}\rangle)].
\]
Since $M(K_{n})=J_{n}-I_{n}$, whenever there is a term like $P[\frac{1}%
{\sqrt{2}}(\left\vert u_{s_{k}+1}\right\rangle |w_{s_{k}^{\prime}}%
\rangle-\left\vert u_{t_{k}+1}\right\rangle |w_{t_{k}^{\prime}}\rangle)]$ in
the sum above, there is a term like $P[\frac{1}{\sqrt{2}}(\left\vert
u_{s_{k}+1}\right\rangle |w_{t_{k}^{\prime}}\rangle-\left\vert u_{t_{k}%
+1}\right\rangle |w_{s_{k}^{\prime}}\rangle)]$. The uniform mixture of these
two terms gives arise to the separable density matrix $\frac{1}{2}P\left[
|u^{+}\rangle|w^{-}\rangle\right]  +\frac{1}{2}P\left[  |u^{-}\rangle
|w^{+}\rangle\right]  $, where $|u^{\pm}\rangle=\frac{1}{\sqrt{2}}\left(
|u_{s_{k}+1}\rangle\pm|u_{t_{k}+1}\rangle\right)  $ and $|w^{\pm}\rangle
=\frac{1}{\sqrt{2}}(|w_{s_{k}^{\prime}}\rangle\pm|w_{t_{k}^{\prime}}\rangle)$.
This shows that $\sigma\left(  K_{n}\right)  $ is separable.
\end{proof}

\begin{remark}
Separability of $\sigma\left(  K_{n}\right)  $ does not depend upon the
labeling of $V(K_{n})$. Given a graph $G$, an isomorphism from $V(G)$ to
$V(G)$ is called \emph{automorphism}. Under composition of maps, the set of
the automorphisms of $G$ form a group, denoted by $Aut(G)$, and called
\emph{automorphism group} of $G$. Note that the separability properties of $G$
are invariant under $Aut(G)$. Since $\sigma(K_{n})$ is separable, and since
the automorphism group of $K_{n}$ is the symmetric group $S_{n}$,
$G\cong\sigma(K_{n})$ is also separable.
\end{remark}

\begin{example}
Consider the graph $K_{4}$. The vertices of $K_{4}$ are denoted by $1,2,3$ and
$4$. We associate to these vertices the orthonormal basis $\{\left\vert
1\right\rangle =\left\vert 1\right\rangle \left\vert 1\right\rangle
,\left\vert 2\right\rangle =\left\vert 1\right\rangle \left\vert
2\right\rangle ,\left\vert 3\right\rangle =\left\vert 2\right\rangle
\left\vert 1\right\rangle ,\left\vert 4\right\rangle =\left\vert
2\right\rangle \left\vert 2\right\rangle \}$. In terms of this basis$\,\sigma
(K_{4})$ can be written as
\begin{multline*}
\sigma(K_{4})=\frac{1}{12}\left[
\begin{array}
[c]{rrrr}%
3 & -1 & -1 & -1\\
-1 & 3 & -1 & -1\\
-1 & -1 & 3 & -1\\
-1 & -1 & -1 & 3
\end{array}
\right]  =\frac{1}{6}P[\left\vert 1\right\rangle \frac{1}{\sqrt{2}}(\left\vert
1\right\rangle -\left\vert 2\right\rangle )]+\frac{1}{6}P[\frac{1}{\sqrt{2}%
}(\left\vert 1\right\rangle -\left\vert 2\right\rangle )\left\vert
1\right\rangle ]\\
+P[\frac{1}{\sqrt{2}}(\left\vert 1\right\rangle -\left\vert 2\right\rangle
)\left\vert 2\right\rangle ]+P[\left\vert 2\right\rangle \frac{1}{\sqrt{2}%
}(\left\vert 1\right\rangle -\left\vert 2\right\rangle )]+\frac{1}{3}%
\{\frac{1}{2}P[\frac{1}{\sqrt{2}}\left(  \left\vert 11\right\rangle
-\left\vert 22\right\rangle \right)  ]+\frac{1}{2}P[\frac{1}{\sqrt{2}}\left(
\left\vert 12\right\rangle -\left\vert 21\right\rangle \right)  ]\}.
\end{multline*}
Each of the first four terms in the above expression is a projector on a
product state, while the last two terms give arise to the separable density
matrix $\frac{1}{2}P\left[  \left\vert -\right\rangle \left\vert
+\right\rangle \right]  +\frac{1}{2}P\left[  \left\vert +\right\rangle
\left\vert -\right\rangle \right]  $, where $\left\vert \pm\right\rangle
\overset{def}{=}\frac{1}{\sqrt{2}}\left(  \left[  \left\vert 1\right\rangle
\pm\left\vert 2\right\rangle \right]  \right)  $. Thus $\sigma(K_{4})$ is
separable in $\mathbb{C}^{2}\otimes\mathbb{C}^{2}$.
\end{example}

\begin{lemma}
\label{complete1}The complete graph on $n>1$ vertices is not a tensor product
of graphs.
\end{lemma}

\begin{proof}
It is clear that, if $n$ is prime, $K_{n}$ is not a tensor product of graphs.
We then assume that $n$ is not a prime. Suppose that there exist graphs $G$
and $H$, respectively on $p$ and $s$ vertices, such that $K_{ps}=G\otimes H$.
Let $\left\vert E(G)\right\vert =q$ and $\left\vert E(H)\right\vert =t$. Then,
by the degree-sum formula, $2q\leq p\left(  p-1\right)  $ and $2t\leq s\left(
s-1\right)  $. So, $2q\cdot2t\leq p\left(  p-1\right)  s\left(  s-1\right)
=ps\left(  ps-p-s+1\right)  $. Now, observe that $\left\vert V(G\otimes
H)\right\vert =ps$ and $\left\vert E(G\otimes H)\right\vert =2\left(
qt\right)  $. Therefore, $G\otimes H=K_{ps}$ if and only if $ps\left(
ps-1\right)  =2\cdot2qt$, which is true if and only if $p=s=1$. This occurs
only when $n=1$.
\end{proof}

\begin{theorem}
\label{cecilia}Given a graph $G\otimes H$, the density matrix $\sigma(G\otimes
H)$ is separable. However if a density matrix $\sigma\left(  L\right)  $ is
separable it does not necessarily mean that $L=G\otimes H$, for some graphs
$G$ and $H$.
\end{theorem}

\begin{proof}
The theorem follows from Lemma \ref{prod} together with Lemma \ref{complete}
and Lemma \ref{complete1}.
\end{proof}

\begin{remark}
Not always is $\sigma\left(  G\right)  \otimes\sigma\left(  G\right)  $ the
density matrix of a graph. However, we observe the following. A \emph{weighted
graph} is a graph with each of its edges labeled by a real number. Let $W$ be
a weighted graph defined as follows: $V=\{ij^{\prime}:i,j^{\prime}=1,2\}$; the
edges of $W$ are
\[
\{11^{\prime},12^{\prime}\},\{11^{\prime},21^{\prime}\},\{12^{\prime
},22^{\prime}\},\{21^{\prime},22^{\prime}\},\{11^{\prime},22^{\prime
}\},\{12^{\prime},21^{\prime}\},
\]
with weights $\frac{1}{2},\frac{1}{2},\frac{1}{2},\frac{1}{2},-\frac{1}{2}$
and $-\frac{1}{2}$, respectively. We write $W:=W(1,2;1^{\prime},2^{\prime})$.
Let $G$ be a graph on the vertices $1,2,...,n$ and the edges $\{i_{_{1}%
},j_{_{1}}\},\{i_{_{2}},j_{_{2}}\},...,\{i_{_{m}},j_{_{m}}\}$, where $1\leq
i_{k},j_{k}\leq n$. Let $H$ be a graph on the vertices $1^{\prime},2^{\prime
},...,p^{\prime}$ and the edges $\{s_{_{1}}^{\prime},t_{_{1}}^{\prime
}\},\{s_{_{2}}^{\prime},t_{_{2}}^{\prime}\},...,\{s_{_{q}}^{\prime},t_{_{q}%
}^{\prime}\}$, then
\[
\sigma\left(  G\right)  \otimes\sigma\left(  G\right)  =\frac{1}{mq}\sum
_{k=1}^{m}\sum_{l=1}^{q}\sigma(i_{k},j_{k};s_{l}^{\prime},t_{l}^{\prime}).
\]

\end{remark}

\subsection{Stars}

A \emph{star graph} (for short, \emph{star}) on $n$ vertices $v_{1}%
,v_{2},...,v_{n}$, denoted by $K_{1,n-1}$, is the graph whose set of edges is
$\{\{v_{1},v_{i}\}:i=2,3,..,n\}$. Quantum dynamics on stars has been studied
in the context of quantum chaos \cite{KMW03}.

\begin{definition}
[Entangled edge]Let $G$ be a graph on $n=pq$ vertices, $v_{1},v_{2},...,v_{n}%
$, The $k$-th edge $\{v_{i_{k}},v_{j_{k}}\}$ of $G$ is identified with the
pure density matrix $P[\frac{1}{\sqrt{2}}\left(  |v_{i_{k}}\rangle-|v_{j_{k}%
}\rangle\right)  ]$, where $\left\vert v_{i_{k}}\right\rangle =\left\vert
u_{s_{k}+1}\right\rangle \left\vert w_{t_{k}}\right\rangle $ and $\left\vert
v_{j_{k}}\right\rangle =|u_{s_{k}^{\prime}+1}\rangle|w_{t_{k}^{\prime}}%
\rangle$, with $i_{k}=s_{k}q+t_{k}$ and $j_{k}=s_{k}^{\prime}q+t_{k}^{\prime}%
$, $0\leq s_{k},s_{k}^{\prime}\leq p-1$ and $1\leq t_{k},t_{k}^{\prime}\leq
q$. The vectors $\left\vert u_{i}\right\rangle $'s and $|w_{j}\rangle$'s form
orthonormal bases of $\mathbb{C}^{p}$ and $\mathbb{C}^{q}$, respectively. The
edge $\{v_{i_{k}},v_{j_{k}}\}$ is said to be \emph{entangled} if $s_{k}\neq
s_{k}^{\prime}$ and $t_{k}\neq t_{k}^{\prime}$.
\end{definition}

\begin{theorem}
\label{star}The density matrix $\sigma(K_{1,n-1})$ is entangled for
$n=pq\geq4$.
\end{theorem}

\begin{proof}
Consider the graph $G=K_{1,n-1}$ on $n=p\cdot q\geq4$ vertices, $v_{1}%
,v_{2},...,v_{n}$. Then%
\[
\sigma\left(  G\right)  =\frac{1}{n-1}%
{\displaystyle\sum\limits_{k=2}^{n}}
\sigma\left(  H_{1k}\right)  =\frac{1}{n-1}%
{\displaystyle\sum\limits_{k=2}^{n}}
P[\frac{1}{\sqrt{2}}\left(  \left\vert v_{1}\right\rangle -\left\vert
v_{k}\right\rangle \right)  ].
\]
We are going to examine separability of $\sigma\left(  G\right)  $ in
$\mathbb{C}_{A}^{p}\otimes\mathbb{C}_{B}^{q}$, where $\mathbb{C}_{A}^{p}$ and
$\mathbb{C}_{B}^{q}$ are associated to two quantum mechanical systems
$\emph{S}_{A}$ and $\emph{S}_{B}$, respectively. Let $\left\{  \left\vert
u_{1}\right\rangle ,\left\vert u_{2}\right\rangle ,...,\left\vert
u_{p}\right\rangle \right\}  $ and $\left\{  \left\vert w_{1}\right\rangle
,\left\vert w_{2}\right\rangle ,...,\left\vert w_{q}\right\rangle \right\}  $
be orthonormal bases of $\mathbb{C}_{A}^{p}$ and $\mathbb{C}_{B}^{q}$,
respectively. So,%
\[
\sigma\left(  G\right)  =\frac{1}{n-1}%
{\displaystyle\sum\limits_{k=2}^{n}}
P\frac{1}{\sqrt{2}}[\left(  \left\vert u_{1}\right\rangle \left\vert
w_{1}\right\rangle -\left\vert u_{s_{k}+1}\right\rangle \left\vert w_{t_{k}%
}\right\rangle \right)  ],
\]
where $k=s_{k}q+t_{k}$, $0\leq s_{k}\leq p-1$ and $1\leq t_{k}\leq q$. Thus%
\begin{align*}
\sigma\left(  G\right)   &  =\frac{1}{n-1}\left\{
{\displaystyle\sum\limits_{j=2}^{q}}
P[\left\vert u_{1}\right\rangle \frac{1}{\sqrt{2}}\left(  \left\vert
w_{1}\right\rangle -\left\vert w_{j}\right\rangle \right)  ]\right. \\
&  +\left.
{\displaystyle\sum\limits_{i=2}^{p}}
P[\frac{1}{\sqrt{2}}\left(  \left\vert u_{1}\right\rangle -\left\vert
u_{i}\right\rangle \right)  \left\vert w_{1}\right\rangle ]+%
{\displaystyle\sum\limits_{j=2}^{q}}
{\displaystyle\sum\limits_{i=2}^{p}}
P[\frac{1}{\sqrt{2}}\left(  \left\vert u_{1}\right\rangle \left\vert
w_{1}\right\rangle -\left\vert u_{i}\right\rangle \left\vert w_{j}%
\right\rangle \right)  ]\right\}  .
\end{align*}
Consider now the following two dimensional projectors: $P=\left\vert
u_{1}\right\rangle \left\langle u_{1}\right\vert +\left\vert u_{2}%
\right\rangle \left\langle u_{2}\right\vert $ and $Q=\left\vert w_{1}%
\right\rangle \left\langle w_{1}\right\vert +\left\vert w_{2}\right\rangle
\left\langle w_{2}\right\vert $. Then%
\begin{align*}
\left(  P\otimes Q\right)  \sigma(G)\left(  P\otimes Q\right)   &  =\frac
{1}{n-1}\left\{  \frac{n-4}{2}P\left[  \left\vert u_{1}\right\rangle
\left\vert w_{1}\right\rangle \right]  +P[\frac{1}{\sqrt{2}}\left(  \left\vert
u_{1}\right\rangle -\left\vert u_{2}\right\rangle \right)  \left\vert
w_{1}\right\rangle ]\right. \\
&  +\left.  P[\left\vert u_{1}\right\rangle \frac{1}{\sqrt{2}}\left(
\left\vert w_{1}\right\rangle -\left\vert w_{2}\right\rangle \right)
]+P[\frac{1}{\sqrt{2}}\left(  \left\vert u_{1}\right\rangle \left\vert
w_{1}\right\rangle -\left\vert u_{2}\right\rangle \left\vert w_{2}%
\right\rangle \right)  ]\right\}  .
\end{align*}
In the basis $\left\{  \left\vert u_{1}\right\rangle \left\vert w_{1}%
\right\rangle ,\left\vert u_{1}\right\rangle \left\vert w_{2}\right\rangle
,\left\vert u_{2}\right\rangle \left\vert w_{1}\right\rangle ,\left\vert
u_{2}\right\rangle \left\vert w_{2}\right\rangle \right\}  $, we have%
\begin{equation}
\left[  \left(  P\otimes Q\right)  \sigma(G)\left(  P\otimes Q\right)
\right]  ^{\intercal_{B}}=\frac{1}{n-1}\left[
\begin{array}
[c]{rrrr}%
\frac{n-1}{2} & -\frac{1}{2} & -\frac{1}{2} & 0\\
-\frac{1}{2} & \frac{1}{2} & -\frac{1}{2} & 0\\
-\frac{1}{2} & -\frac{1}{2} & \frac{1}{2} & 0\\
0 & 0 & 0 & \frac{1}{2}%
\end{array}
\right]  . \label{something}%
\end{equation}
The eigenvalues of the above matrix are $\{[\frac{1}{2\left(  n-1\right)
}],[\frac{1}{n-1}],[\frac{1}{4}(1\pm\sqrt{(n-1)^{2}+8}/\left(  n-1\right)
)]\}$. As $n\geq4$, $\frac{1}{4}(1-\sqrt{(n-1)^{2}+8}/(n-1))<0$. Hence, by
Criterion \ref{pht}, the matrix $\left(  P\otimes Q\right)  \sigma(G)\left(
P\otimes Q\right)  $ is entangled and then also $\sigma(G)$ is entangled (Note
that this matrix is not normalized.)
\end{proof}

\begin{example}
Consider the graph $G=\left(  \left\{  1,2,3,4\right\}  ,\left\{  \left\{
1,2\right\}  ,\left\{  1,3\right\}  ,\left\{  1,4\right\}  ,\left\{
2,3\right\}  \right\}  \right)  $. We test the separability of $\sigma(G)$ in
$\mathbb{C}_{A}^{2}\otimes\mathbb{C}_{B}^{2}$, with respect to the orthonormal
basis $\left\{  \left\vert 1\right\rangle |1\rangle,\left\vert 1\right\rangle
\left\vert 2\right\rangle ,\left\vert 2\right\rangle \left\vert 1\right\rangle
,\left\vert 2\right\rangle \left\vert 2\right\rangle \right\}  $. In this
basis,
\[
\sigma(G)=\frac{1}{8}\left[
\begin{array}
[c]{rrrr}%
3 & -1 & -1 & -1\\
-1 & 2 & -1 & 0\\
-1 & -1 & 2 & 0\\
-1 & 0 & 0 & 1
\end{array}
\right]  .
\]
It can be easily verified that $\left(  \sigma(G)\right)  ^{T_{B}}=\sigma(G)$.
As a consequence, all the eigenvalues of $\left(  \sigma(G)\right)  ^{T_{B}}$
are nonnegative, as $\sigma(G)$ is positive semidefinite. It follows from
Criterion \ref{pht} that $\sigma(G)$ is separable in $\mathbb{C}^{2}%
\otimes\mathbb{C}^{2}$. Consider now the star $K_{1,3}=\left(  \left\{
1,2,3,4\right\}  ,\left\{  \left\{  1,2\right\}  ,\left\{  1,3\right\}
,\left\{  1,4\right\}  \right\}  \right)  $. Observe that $K_{1,3}$ is
obtained from $G$ with the removal of the edge $\left\{  2,3\right\}  $. With
respect to the above mentioned basis, we have%
\[%
\begin{tabular}
[c]{lll}%
$\sigma(K_{1,3})=\frac{1}{6}\left[
\begin{array}
[c]{rrrr}%
3 & -1 & -1 & -1\\
-1 & 1 & 0 & 0\\
-1 & 0 & 1 & 0\\
-1 & 0 & 0 & 1
\end{array}
\right]  $ & and & $\left(  \sigma(K_{1,3})\right)  ^{T_{B}}=\frac{1}%
{6}\left[
\begin{array}
[c]{rrrr}%
3 & -1 & -1 & 0\\
-1 & 1 & -1 & 0\\
-1 & -1 & 1 & 0\\
0 & 0 & 0 & 1
\end{array}
\right]  .$%
\end{tabular}
\
\]
The eigenvalues of $\left(  \sigma(K_{1,3})\right)  ^{T_{B}}$ are $\frac{1}%
{6},\frac{1}{3},\frac{1}{12}\sqrt{17}+\frac{1}{4}$ and $\frac{1}{4}-\frac
{1}{12}\sqrt{17}$. It follows from the Criterion \ref{pht} that $\sigma
(K_{1,3})$ is entangled in $\mathbb{C}^{2}\otimes\mathbb{C}^{2}$.
\end{example}

\begin{remark}
A density matrix $\rho_{AB}$ acting on $\mathbb{C}_{A}^{p}\otimes
\mathbb{C}_{B}^{q}$ is said to be \emph{distillable} if there exist a positive
integer $k$ and two $2$-dimensional projectors $P_{1}:\left(  \mathbb{C}%
_{A}^{p}\right)  ^{\otimes k}\longrightarrow\mathbb{C}^{2}$ and $P_{2}:\left(
\mathbb{C}_{B}^{q}\right)  ^{\otimes k}\longrightarrow\mathbb{C}^{2}$ such
that $\left(  \left(  P_{1}\otimes P_{2}\right)  \rho_{AB}^{\otimes k}\left(
P_{1}\otimes P_{2}\right)  \right)  ^{\intercal_{B}}\ngeqq0$. An entangled
density matrix which is not distillable is called \emph{bound entangled}.
Theorem \ref{star} actually shows that not only $\sigma(K_{1,n-1})$ is
entangled but also distillable in $\mathbb{C}^{p}\otimes\mathbb{C}^{q}$ where
$n=pq\geq4$.
\end{remark}

\begin{proof}
[Second proof of Theorem \ref{star}]Let $G$ be a graph on $n$ vertices and $m$
edges. Suppose that $G$ has $l_{i} $ loops at the vertex $v_{i}$. Then
$|E(G)|=m+\sum_{i=1}^{n}l_{i}$ edges. We associate to $G$ the following
density matrix%
\begin{equation}
\sigma_{\circ}(G)\overset{def}{=}\left(  2m+%
{\displaystyle\sum\limits_{i=1}^{n}}
l_{i}\right)  ^{-1}(\Delta(G)-M(G))+\left(  2m+%
{\displaystyle\sum\limits_{i=1}^{n}}
l_{i}\right)  ^{-1}\Delta_{\circ}(G),\text{ where }\Delta_{\circ}(G)=%
{\displaystyle\bigoplus\limits_{i=1}^{n}}
l_{i}P[|v_{i}\rangle]. \label{newlap}%
\end{equation}
The matrix $\sigma_{\circ}(G)$ generalizes the notion of density matrix of a
graph by taking into account loops. The star $K_{1,3}$ is called \emph{claw}.
We denote by $K_{1,3}^{+l}$ a claw with $l$ loops at the vertex of degree $3$.
In order to prove the theorem, we first show that $K_{1,3}^{+l}$ is entangled
in $\mathbb{C}_{A}^{2}\otimes\mathbb{C}_{B}^{2}$. The vertices of $K_{1,3}$
are denoted by $u_{1}w_{1},u_{1}w_{2},u_{2}w_{1},u_{2}w_{2}$, where
$d(u_{1}w_{1})=3$. We then have%
\begin{align*}
\sigma_{\circ}(K_{1,3}^{+l})  &  =\frac{2}{l+6}\left\{  \frac{l}%
{2}P[\left\vert u_{1}\right\rangle \left\vert w_{1}\right\rangle ]+P[\frac
{1}{\sqrt{2}}(\left\vert u_{1}\right\rangle \left\vert w_{1}\right\rangle
-\left\vert u_{2}\right\rangle \left\vert w_{2}\right\rangle )]\right. \\
&  +\left.  P[\frac{1}{\sqrt{2}}(\left\vert u_{1}\right\rangle -\left\vert
u_{2}\right\rangle \left\vert w_{1}\right\rangle ]+P[\left\vert u_{1}%
\right\rangle \frac{1}{\sqrt{2}}(\left\vert w_{1}\right\rangle -\left\vert
w_{2}\right\rangle )]\right\}  .
\end{align*}
One can check that $[\sigma_{\circ}(K_{1,3}^{+l})]^{\intercal_{B}}\ngeqslant
0$. By Theorem \ref{pht}, $\sigma_{\circ}(K_{1,3}^{+l})$ is entangled in
$\mathbb{C}_{A}^{2}\otimes\mathbb{C}_{B}^{2}$, for all $l\geq0$. Let $A$ be
the matrix in Equation \ref{something}. Taking $n-4=l$, $A=\frac{l+6}{2\left(
l+3\right)  }\sigma_{\circ}(K_{1,3}^{+l})$. Then $A$ is entangled in
$\mathbb{C}_{A}^{2}\otimes\mathbb{C}_{B}^{2}$. This shows that $\sigma
(K_{1,n-1})$ is entangled in $\mathbb{C}_{A}^{p}\otimes\mathbb{C}_{B}^{q}$.
\end{proof}

\begin{proposition}
Separability of $\sigma(K_{1,n-1})$, with $n=pq\geq4$, does not depend on the
labeling of $V(K_{1,n-1})$.
\end{proposition}

\begin{proof}
In $K_{1,n-1}$, the vertex of degree $\left(  n-1\right)  $ is called
\emph{root}, the other vertices are called \emph{leafs}. We define two types
of isomorphisms for stars: \emph{Leaf-shuffling}) An isomorphism $\iota$
acting on $V(K_{1,n-1})$ is called a \emph{leaf-shuffling} if $\iota(r)=r$,
where $r$ is the root of $K_{1,n-1}$; \emph{Root-swapping}) An isomorphism
$\iota$ acting on $V(K_{1,n-1})$ is called a \emph{root-swapping} if
$\iota(r)=v$, where $r$ is the root of $K_{1,n-1}$ and $v$ is a leaf. All
graphs in the isomorphism class of $K_{1,n-1}$ can be obtained by combining
leaf-shuffling and root-swapping. It is clear that leaf-shuffling is an
automorphism and hence it does not change the separability property of
$\sigma(K_{1,n-1})$. We now prove that this is the case also for
root-swapping. We label the vertices of a graph $G\cong K_{1,n-1}$ as
$u_{1}w_{1},u_{1}w_{2},...,u_{1}w_{q},u_{2}w_{1},u_{2}w_{2},...,u_{2}%
w_{q},...,u_{p}w_{q} $. Let $u_{1}w_{1}$ be the root of $G$ and let
$\iota:V(G)\longrightarrow V(H)$ be a root-swapping. Then, the root of $H$ is
$\iota(u_{1}w_{1})=u_{i}w_{j}$, where $\left(  1,1\right)  \neq\left(
i,j\right)  $. Denote by $P_{1\leftrightarrow i}$ and $Q_{1\leftrightarrow j}$
the permutation matrices defined as follows: $P_{1\leftrightarrow i}%
|u_{1}\rangle=|u_{i}\rangle$ and $P_{1\leftrightarrow i}|u_{i^{\prime}}%
\rangle=|u_{i^{\prime}}\rangle$ for $i^{\prime}\neq1$; $Q_{1\leftrightarrow
j}|w_{1}\rangle=|w_{j}\rangle$ and $Q_{1\leftrightarrow j}|w_{j^{\prime}%
}\rangle=|w_{j^{\prime}}\rangle$ for $j^{\prime}\neq1$. Then we have $\left(
P_{1\leftrightarrow i}\otimes Q_{1\leftrightarrow j}\right)  \sigma(G)\left(
P_{1\leftrightarrow i}\otimes Q_{1\leftrightarrow j}\right)  ^{\intercal
}=\sigma(H)$. Then $\sigma(G)$ is entangled if and only if $\sigma(H)$ is entangled.
\end{proof}

\begin{remark}
For $K_{1,n-1}$, with $n=pq\geq4$, $K_{1,n-1}\ncong G\otimes H$, where
$|V(G)|=p$ and $|V(H)|=q$.
\end{remark}

\subsection{Perfect matchings}

A \emph{matching} of a graph is a set of vertex-disjoint edges. A
\emph{perfect matching} of a graph $G$ is a matching spanning $V(G)$.

\begin{definition}
[e-matching; pe-matching]An \emph{e-matching} is a matching having all edges
entangled. Each vertex of an e-matching on $n=pq$ vertices can be labeled by
an ordered pairs $(i,j) $, where $1\leq i\leq p$ and $1\leq j\leq q$. A
\emph{pe-matching} of a graph $G$ is an e-matching spanning $V(G)$.
\end{definition}

\begin{theorem}
\label{mes}Let $G$ be a graph on $n=2p$ vertices. If all the entangled edges
of $G$ belong to the same pe-matching then $\sigma(G)$ is separable in
$\mathbb{C}^{2}\otimes\mathbb{C}^{p}$.
\end{theorem}

Our proof of the theorem involves the use of the following concepts.

\begin{definition}
[Criss-cross]A \emph{criss-cross} is a set
$\{\{(k,i),(l,j)\},\{(k,j),(l,i)\}\}$ of two edges belonging to an e-matching
on $n=pq$ vertices.
\end{definition}

\begin{definition}
[Tally-mark]A set
\[
\{(k,i_{1}),(l,i_{2})\},\{(k,i_{2}),(l,i_{3})\},...,\{(k,i_{s+1}%
),(l,i_{s+2})\},\{(k,i_{s+2}),(l,i_{1})\}
\]
of $s+2$ edges, where $k<l$, $s\geq0$ and $i_{1}<i_{2}<\cdots<i_{s+2}$,
belonging to an e-matching on $n=pq$ vertices, is called a \emph{tally-mark}.
(Note that a criss-cross is a tally-mark with two edges.)
\end{definition}

\begin{definition}
[Canonical pe-matching]Let $H$ be an e-matching on $n=pq$ vertices. Then $H$
is said to be \emph{canonical }if $H=H_{1}\uplus H_{2}\uplus\cdots\uplus
H_{k}$, where every graph $H_{1},H_{2},...,H_{k}$ is a tally-mark.
\end{definition}

\begin{lemma}
\label{cpe}From any pe-matching $H$ on $n=2p$ vertices, labeled by $(i,j)$,
where $i=1,2$ and $1\leq j\leq p$, we can always obtain a canonical
pe-matching by applying a permutation to the second label of all the vertices
of $H$.
\end{lemma}

\begin{proof}
Let $H$ be a pe-matching as in the statement of the lemma. Any pe-matching can
be taken as a set of criss-crosses and e-matchings, the latter being of the forms:

\begin{itemize}
\item $H_{1}$ such that $E(H_{1})=\{\{(1,i_{1}),(2,j_{1})\},\{(1,i_{2}%
),(2,j_{2})\},...,\{(1,i_{k}),(2,j_{k})\}\}$, where $\{i_{1},i_{2}%
,...,i_{k}\}=\{j_{1},j_{2},...,j_{k}\}$;

\item $H_{2}$ such that $E(H_{2})=\{\{(1,i_{1}^{\prime}),(2,j_{1}^{\prime
})\},\{(1,i_{2}^{\prime}),(2,j_{2}^{\prime})\},...,\{(1,i_{l}^{\prime
}),(2,j_{l}^{\prime})\}\}$, where $\{i_{1}^{\prime},i_{2}^{\prime}%
,...,i_{l}^{\prime}\}=\{j_{1}^{\prime},j_{2}^{\prime},...,j_{l}^{\prime}\}$
and $\{i_{1},i_{2},...,i_{k}\}\cap\{i_{1}^{\prime},i_{2}^{\prime}%
,...,i_{l}^{\prime}\}=\emptyset$;

\item $...$
\end{itemize}

We describe an algorithm to obtain a tally-mark from any of the above
e-matchings. It is sufficient to describe the algorithm for $H_{1}$. Without
loss of generality we take $i_{1}<i_{2}<\cdots<i_{k}$. We permute the $2^{nd}$
labels of the edges of $H_{1}$, to form one or more disjoint tally-marks.
Consider the $s$-th step in the construction: if $j_{s}<i_{s}$, we have
completed a tally-mark; if $j_{s}>i_{s}$ then we perform a permutation on the
$2^{nd}$ label of a vertex $(\cdot,i)$ acting on indices $i>i_{s}$, which maps
the edge $\{(1,i_{s}),(2,j_{s})\}$ to the edge $\{(1,i_{s}),(2,i_{s+1})\}$
(adding another downstroke to a tally-mark, yet incomplete). It is easy to see
that applying this rule successively to the labels $\{i_{1},i_{2},..,i_{k}\}$,
in ascending order, produces a set of one or more disjoint tally-marks.
\end{proof}

\begin{example}
In Figure 1, a pe-matching (top graph) is transformed in a canonical
pe-matching by applying a permutation on the second labels of the vertices. We
first apply the permutation $(2$ $3)$ (central graph). We then apply the
permutation $(3$ $5)$ (bottom graph).
\begin{center}
\includegraphics[
height=2.3436in,
width=2.8643in
]%
{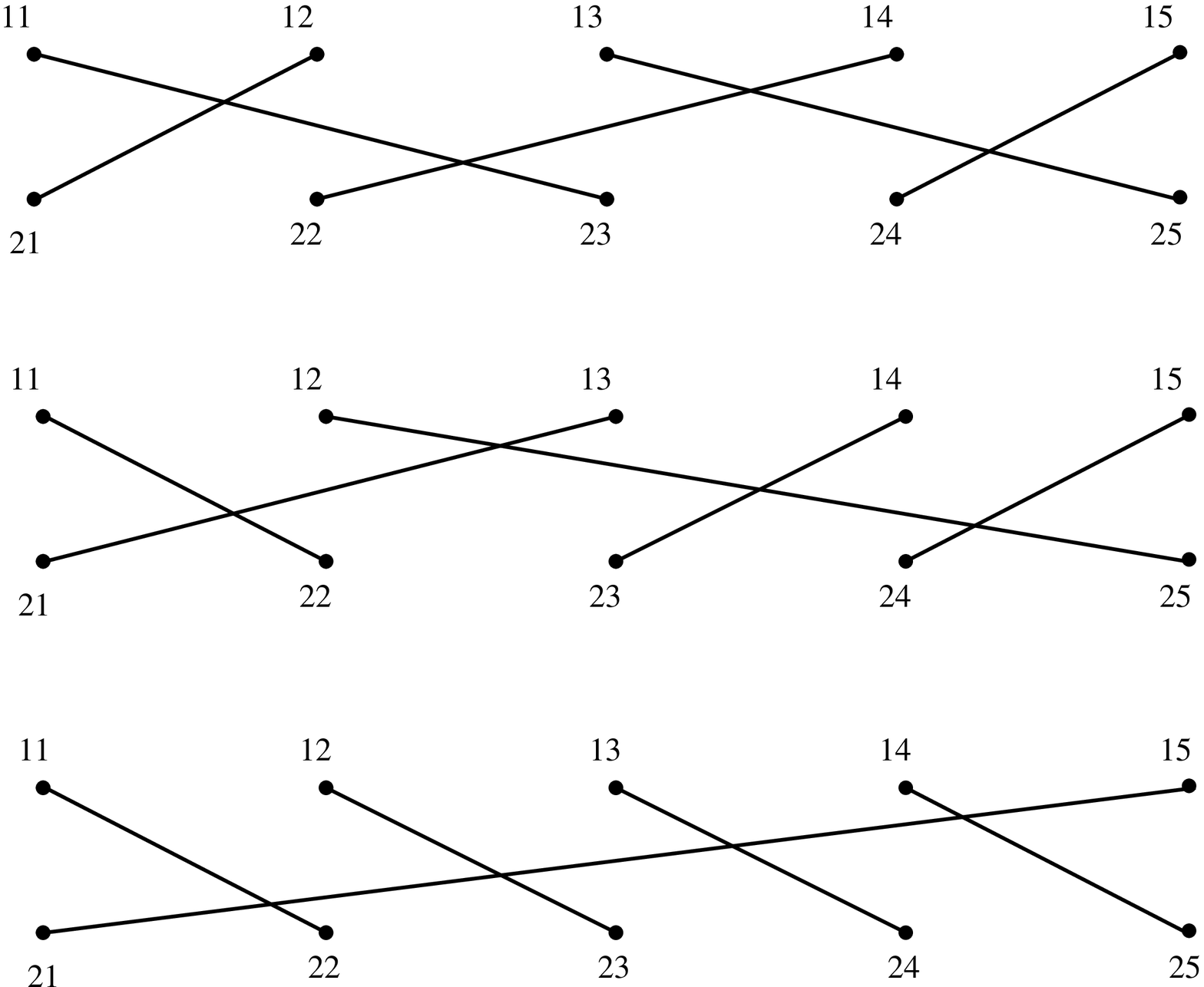}%
\\
Figure 1
\end{center}

\end{example}

\begin{lemma}
\label{cpe2}Let $H$ be a tally-mark on $n=2k+2$ vertices. Then $\sigma(H)$ is
separable in $\mathbb{C}^{2}\otimes\mathbb{C}^{k+1}$.
\end{lemma}

\begin{proof}
Let $H$ be a tally-mark. Let us assume that $H$ is not a criss-cross. In fact,
if $H$ is a criss-cross then $\sigma(H)$ is obviously separable in
$\mathbb{C}^{2}\otimes\mathbb{C}^{2}$. Let
\[
E(H)=\{\{(1,i_{0}),(2,i_{1})\},\{(1,i_{1}),(2,i_{2})\},...,\{(1,i_{k-1}%
),(2,i_{k})\},\{(1,i_{k}),(2,i_{0})\}\},
\]
where $i_{0}<i_{1}<\cdots<i_{k}$. We associate the vector $|l\rangle
|i_{s}\rangle$ to the vertex $(l,i_{s})\in V(H)$, where $l=1,2$ and
$s=0,1,2,...,k$. Then
\[
\sigma(H)=\frac{1}{k+1}%
{\displaystyle\sum\limits_{s=0}^{k}}
P[\frac{1}{\sqrt{2}}(|1\rangle|i_{s}\rangle-|2\rangle|i_{\left(  s+1\right)
\operatorname{mod}\left(  k+1\right)  }\rangle].
\]
Let us consider the permutation $g$ on the $\left(  k+1\right)  $ letters
$i_{0},i_{1},...,i_{k}$ defined as follows: $g:i_{s}\longmapsto i_{\left(
s+1\right)  \operatorname{mod}\left(  k+1\right)  }$, where $0\leq s\leq k$.
The order of $g$ is then $\left(  k+1\right)  $. Let $\Gamma=\left\langle
g\right\rangle \cong\mathbb{Z}_{k+1}$ and $\rho_{reg}\left(  g\right)  =\Pi$.
One can check that
\begin{equation}
\sigma(H)=\frac{1}{k+1}%
{\displaystyle\sum\limits_{s=0}^{k}}
\left(  I_{2}\otimes\Pi^{s}\right)  P[\frac{1}{\sqrt{2}}(|1\rangle
|i_{0}\rangle-|2\rangle|i_{1}\rangle]\left(  I_{2}\otimes\Pi^{k+1-s}\right)  ,
\label{tally}%
\end{equation}
where $I_{2}$ acts on the Hilbert space spanned by the vectors $|1\rangle$ and
$|2\rangle$. We are now looking for the density matrices acting on
$\mathbb{C}^{2}\otimes\mathbb{C}^{k+1}$, which remains invariant under the
action of $\Gamma$. Let%
\[%
\begin{tabular}
[c]{lll}%
$|\psi_{m}\rangle=\frac{1}{\sqrt{k+1}}%
{\displaystyle\sum\limits_{s=0}^{k}}
\exp[\frac{2\pi ism}{k+1}]|i_{s}\rangle,$ &  & where $m=0,1,2,...,k.$%
\end{tabular}
\
\]
Observe that the vectors $|\psi_{m}\rangle$'s are pairwise orthonormal. Then
$\Pi^{l}|\psi_{m}\rangle=\exp[\frac{2\pi ilm}{k+1}]|\psi_{m}\rangle$, for
$l=0,1,2,...,k$ and $m=0,1,2,...,k$. It follows that the $|\psi_{m}\rangle$'s
are eigenvectors of $\Pi^{l}$. Let
\[%
\begin{tabular}
[c]{lll}%
$|\Psi\rangle=%
{\displaystyle\sum\limits_{m=0}^{k}}
(\alpha_{m}|1\rangle+\beta_{m}|2\rangle)|\psi_{m}\rangle$ & where & $%
{\displaystyle\sum\limits_{m=0}^{k}}
(|\alpha_{m}|^{2}+|\beta_{m}|^{2})=1,$%
\end{tabular}
\
\]
be a vector in $\mathbb{C}^{2}\otimes\mathbb{C}^{k+1}$. Then $\left(
I_{2}\otimes\Pi^{l}\right)  |\Psi\rangle\langle\Psi|\left(  I_{2}\otimes
\Pi^{k+1-l}\right)  =|\Psi\rangle\langle\Psi|$, where $l=0,1,2,...,k$, if and
only if $|\Psi\rangle$ is one of the forms $(\alpha_{m}|1\rangle+\beta
_{m}|2\rangle)|\psi_{m}\rangle$, for $m=0,1,2,...,k$. This shows that, for any
density matrix $\rho$ acting on $\mathbb{C}^{2}\otimes\mathbb{C}^{k+1}$, the
following density matrix%
\[
\rho^{\prime}=\frac{1}{k+1}%
{\displaystyle\sum\limits_{s=0}^{k}}
\left(  I_{2}\otimes\Pi^{s}\right)  \rho\left(  I_{2}\otimes\Pi^{k+1-s}%
\right)
\]
is a mixture of all the projectors $P[\frac{1}{\sqrt{2}}(\alpha_{m}%
|1\rangle+\beta_{m}|2\rangle)|\psi_{m}\rangle]$, where $m=0,1,2,...,k$. Hence
$\rho^{\prime}$ is separable. By Equation \ref{tally}, $\sigma(H)$ is also
separable:%
\[
\sigma(H)=\frac{1}{k+1}%
{\displaystyle\sum\limits_{m=0}^{k}}
P[\frac{1}{\sqrt{2}}(|1\rangle-\exp[-\tfrac{2\pi im}{k+1}]|2\rangle)|\psi
_{m}\rangle].
\]

\end{proof}

\bigskip

Given a graph $G$ and a factor $H$ of $G$, we denote by $G\backslash H$ the
graph with adjacency matrix $M(G\backslash H)\overset{def}{=}M(G)-M(H)$.

\bigskip

\begin{proof}
[Proof of Theorem \ref{mes}]Let $G$ be as in the statement of the theorem. In
addition, we assume that $\left\vert E(G)\right\vert =m$. Let $H$ be the
pe-matching containing all the entangled edges of $G$. Then $\sigma
(G)=\frac{p}{m}\sigma(H)+\frac{m-p}{m}\sigma(G\backslash H)$. The density
matrix $\sigma(G\backslash H)$ is separable by assumption. Lemma \ref{cpe}
together with Lemma \ref{cpe2} shows that $\sigma(H)$ is separable in
$\mathbb{C}^{2}\otimes\mathbb{C}^{p}$. This proves the theorem.
\end{proof}

\begin{theorem}
The pe-matching in Figure 2 is separable in $\mathbb{C}^{3}\otimes
\mathbb{C}^{4}$.
\begin{center}
\includegraphics[
height=0.9383in,
width=2.1837in
]%
{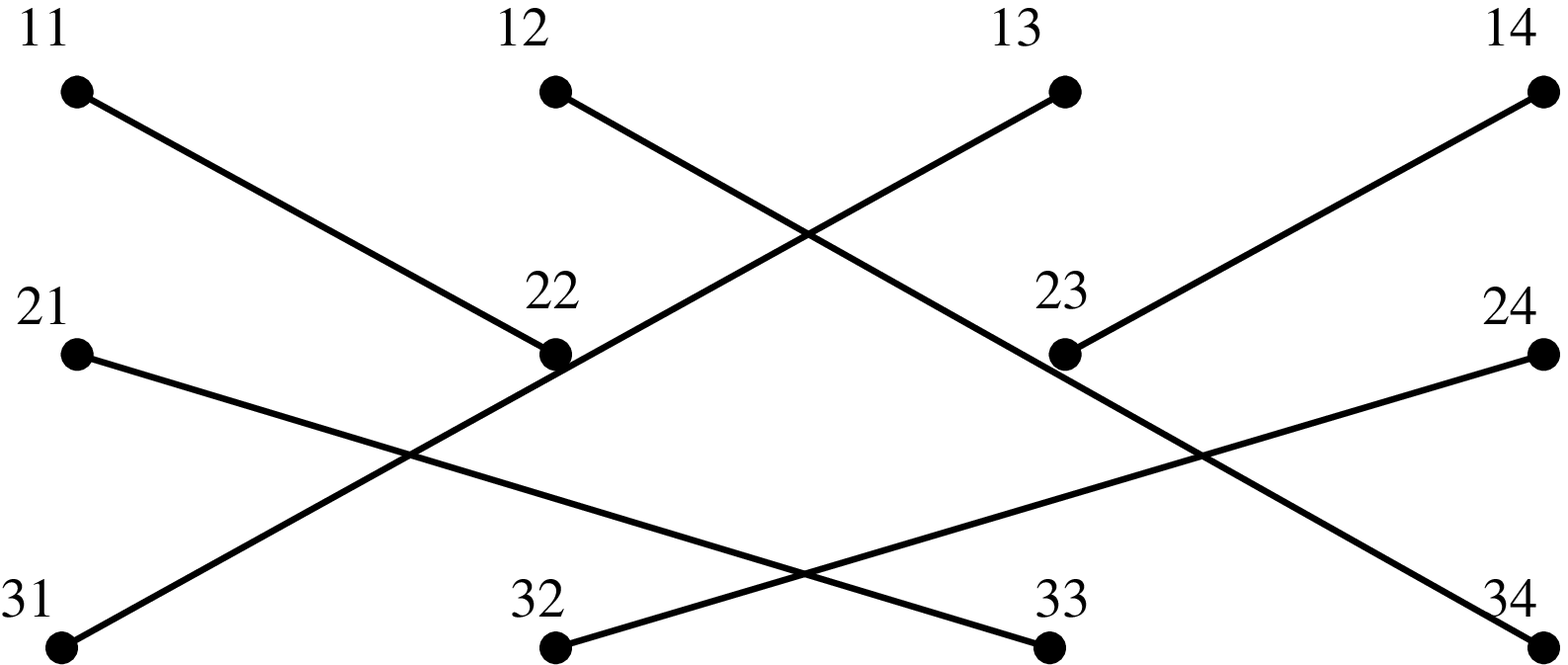}%
\\
Figure 2
\end{center}

\end{theorem}

\begin{proof}
Let $G$ be the pe-matching in the figure. Then $\sigma(G)=\frac{1}{6}%
\sum_{i=1}^{6}P[|\psi_{i}^{-}\rangle]$, where $|\psi_{1}^{\pm}\rangle=\frac
{1}{\sqrt{2}}(|1\rangle|1\rangle\pm|2\rangle|2\rangle)$, $|\psi_{2}^{\pm
}\rangle=\frac{1}{\sqrt{2}}(|1\rangle|4\rangle\pm|2\rangle|3\rangle)$,
$|\psi_{3}^{\pm}\rangle=\frac{1}{\sqrt{2}}(|1\rangle|2\rangle\pm
|3\rangle|4\rangle)$, $|\psi_{4}^{\pm}\rangle=\frac{1}{\sqrt{2}}%
(|1\rangle|3\rangle\pm|3\rangle|1\rangle)$, $|\psi_{5}^{\pm}\rangle=\frac
{1}{\sqrt{2}}(|2\rangle|1\rangle\pm|3\rangle|3\rangle)$ and $|\psi_{6}^{\pm
}\rangle=\frac{1}{\sqrt{2}}(|2\rangle|4\rangle\pm|3\rangle|2\rangle)$. Here
$(\sigma(G))^{\intercal_{B}}=(I_{3}\otimes P)\sigma(G)(I_{3}\otimes
P^{\intercal})$, where
\[
P=\left[
\begin{array}
[c]{cccc}%
0 & 1 & 0 & 0\\
1 & 0 & 0 & 0\\
0 & 0 & 0 & 1\\
0 & 0 & 1 & 0
\end{array}
\right]  .
\]
Then $(\sigma(G))^{\intercal_{B}}\geq0$. Since a density matrix having
positive partial transpose is either separable or bound entangled
\cite{HHH98}, this holds for $\sigma(G)$. We are now going to show that
$\sigma(G)$ is separable in $\mathbb{C}^{3}\otimes\mathbb{C}^{4}$. Let
$|\chi\rangle$ be in the support of $\sigma(G)$. Then $|\chi\rangle=\sum
_{i=1}^{6}a_{i}|\psi_{i}^{-}\rangle=\frac{1}{\sqrt{2}}|1\rangle(a_{1}%
|1\rangle+a_{3}|2\rangle+a_{4}|3\rangle+a_{2}|4\rangle)+\frac{1}{\sqrt{2}%
}|2\rangle(a_{5}|1\rangle-a_{1}|2\rangle-a_{2}|3\rangle+a_{6}|4\rangle
)-\frac{1}{\sqrt{2}}|3\rangle(a_{4}|1\rangle+a_{6}|2\rangle+a_{5}%
|3\rangle+a_{3}|4\rangle)$. So, $|\chi\rangle$ is separable if and only if
$(a_{1},a_{3},a_{4},a_{2})=\lambda(a_{5},-a_{1},-a_{2},a_{6})=\mu(a_{4}%
,a_{6},a_{5},a_{3})$, where $\lambda,\mu\in\mathbb{C}$. Then
\begin{equation}
a_{1}=\lambda a_{5}=\mu a_{4},\text{ }a_{3}=-\lambda a_{1}=\mu a_{6},\text{
}a_{4}=-\lambda a_{2}=\mu a_{5},\text{ }a_{2}=\lambda a_{6}=\mu a_{3}
\label{lamba}%
\end{equation}
Here $\lambda\neq0$. In fact, if $\lambda=0$ and $\mu\neq0$, then $a_{i}=0$
for $i=1,2,...,6$, which is impossible. On the other hand, if $\lambda=\mu=0$
then $a_{1},a_{2},a_{3},a_{4}=0$. Then $|\chi\rangle=\frac{1}{\sqrt{2}%
}|2\rangle(a_{5}|1\rangle+a_{6}|4\rangle)-\frac{1}{\sqrt{2}}|3\rangle
(a_{6}|2\rangle+a_{5}|3\rangle)$, which is entangled as $|a_{5}|^{2}%
+|a_{6}|^{2}\neq0$. Similarly, it can be shown that $\mu\neq0$. Therefore,
from Equation \ref{lamba}, $\lambda^{3}=1$ and $\mu^{2}=\lambda$, and we can
distinguish the following cases.

\noindent\textbf{Case 1.} $(\lambda=\mu=1)$ We have $a_{2}=a_{3}=a_{6}=-a_{1}$
and $a_{4}=a_{5}=a_{1}$. So $|\chi\rangle=\frac{a_{1}}{\sqrt{2}}%
(|1\rangle+|2\rangle-|3\rangle)(|1\rangle-|2\rangle+|3\rangle-|4\rangle)$.

\noindent\textbf{Case 2. }$(\lambda=1,\mu=-1)$ We have $a_{2}=a_{5}%
=a_{6}=a_{1}$ and $a_{3}=a_{4}=-a_{1}$. So $|\chi\rangle=\frac{a_{1}}{\sqrt
{2}}(|1\rangle+|2\rangle+|3\rangle)(|1\rangle-|2\rangle-|3\rangle+|4\rangle)$.

\noindent\textbf{Case 3. }$(\lambda=\omega=e^{2\pi i/3},\mu=-\omega^{2}) $ We
have $a_{2}=a_{1}$, $a_{3}=a_{4}=-\omega a_{1}$, $a_{5}=a_{6}=\omega^{2}a_{1}%
$. So $|\chi\rangle=\frac{a_{1}}{\sqrt{2}}(|1\rangle+\omega^{2}|2\rangle
+\omega|3\rangle)(|1\rangle-\omega|2\rangle-\omega|3\rangle+|4\rangle)$.

\noindent\textbf{Case 4. }$(\lambda=\omega,\mu=\omega^{2})$ We have
$a_{2}=-a_{1}$, $a_{3}=-a_{4}=-\omega a_{1}$, $a_{5}=-a_{6}=\omega^{2}a_{1}$.
So $|\chi\rangle=\frac{a_{1}}{\sqrt{2}}(|1\rangle+\omega^{2}|2\rangle
-\omega|3\rangle)(|1\rangle-\omega|2\rangle+\omega|3\rangle-|4\rangle)$.

\noindent\textbf{Case 5. }$(\lambda=\omega^{2},\mu=\omega)$ We have
$a_{2}=-a_{1}$, $a_{3}=-a_{4}=-\omega^{2}a_{1}$, $a_{5}=-a_{6}=\omega a_{1}$.
So $|\chi\rangle=\frac{a_{1}}{\sqrt{2}}(|1\rangle+\omega|2\rangle-\omega
^{2}|3\rangle)(|1\rangle-\omega^{2}|2\rangle+\omega^{2}|3\rangle-|4\rangle)$.

\noindent\textbf{Case 6. }$(\lambda=\omega^{2},\mu=-\omega)$ We have
$a_{2}=a_{1}$, $a_{3}=a_{4}=-\omega^{2}a_{1}$, $a_{5}=a_{6}=\omega a_{1}$. So
$|\chi\rangle=\frac{a_{1}}{\sqrt{2}}(|1\rangle+\omega|2\rangle+\omega
^{2}|3\rangle)(|1\rangle-\omega^{2}|2\rangle-\omega^{2}|3\rangle+|4\rangle)$.

Thus we can observe that the range of the rank six density matrix $\sigma(G) $
contains only the following six separable states: $|\chi_{1}\rangle=\frac
{1}{\sqrt{3}}(|1\rangle+|2\rangle-|3\rangle)\frac{1}{2}(|1\rangle
-|2\rangle+|3\rangle-|4\rangle)$, $|\chi_{2}\rangle=\frac{1}{\sqrt{3}%
}(|1\rangle+|2\rangle+|3\rangle)\frac{1}{2}(|1\rangle-|2\rangle-|3\rangle
+|4\rangle)$, $|\chi_{3}\rangle=\frac{1}{\sqrt{3}}(|1\rangle+\omega
^{2}|2\rangle+\omega|3\rangle)\frac{1}{2}(|1\rangle-\omega|2\rangle
-\omega|3\rangle+|4\rangle)$, $|\chi\rangle=\frac{1}{\sqrt{3}}(|1\rangle
+\omega^{2}|2\rangle-\omega|3\rangle)\frac{1}{2}(|1\rangle-\omega
|2\rangle+\omega|3\rangle-|4\rangle)$, $|\chi_{5}\rangle=\frac{1}{\sqrt{3}%
}(|1\rangle+\omega|2\rangle-\omega^{2}|3\rangle)\frac{1}{2}(|1\rangle
-\omega^{2}|2\rangle+\omega^{2}|3\rangle-|4\rangle)$ and $|\chi_{6}%
\rangle=\frac{1}{\sqrt{3}}(|1\rangle+\omega|2\rangle+\omega^{2}|3\rangle
)\frac{1}{2}(|1\rangle-\omega^{2}|2\rangle-\omega^{2}|3\rangle+|4\rangle)$.
These states are pairwise orthogonal. As $\sigma(G)$ is proportional to a six
dimensional projector, we can write $\sigma(G)=\frac{1}{6}\sum_{i=1}^{6}%
|\chi_{i}\rangle\langle\chi_{i}|$, and hence $\sigma(G)$ is separable.
\end{proof}

\begin{remark}
The pe-matching $G$ in Figure 3 is entangled in $\mathbb{C}^{3}\otimes
\mathbb{C}^{4}$. In fact, it can be shown that $(\sigma(G))^{\intercal_{B}%
}\ngeqslant0$.
\begin{center}
\includegraphics[
height=0.9383in,
width=2.1975in
]%
{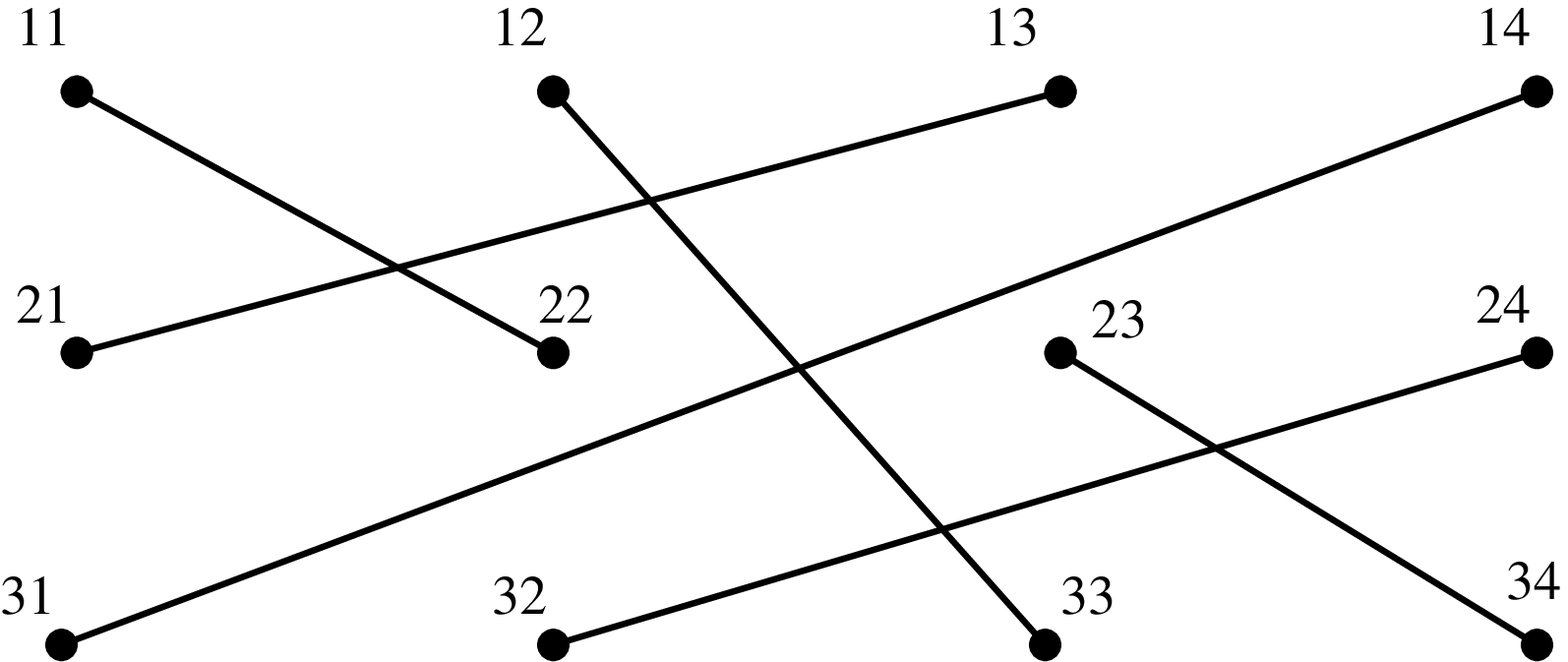}%
\\
Figure 3
\end{center}

\end{remark}

\subsection{The Petersen graph}

Let $v,k$ and $i$ be fixed positive integers, with $v\geq k\geq i$. Let $S$ be
an $n$-elements set. The \emph{Johnson graph} $J(v,k,i)$ is defined as
follows: the vertices of $J(v,k,i)$ are the $k$-elements subsets of $S$; two
vertices are adjacent if their intersection has size $i$. The graph $J(5,2,0)$
is called \emph{Petersen graph} and it has a number of important properties.
For example, it is strongly-regular and transitive. A graph $G$ that is not
complete is said to be \emph{strongly-regular} if it is regular, every pair of
adjacent vertices has a the same number of common neighbours, and every pair
of nonadjacent vertices has the same number of common neighbours. A graph $G$
is said to be \emph{transitive} if $Aut(G)$ acts transitively on $V(G)$. A
permutation group $\Gamma$ \emph{acts transitively} on a set $S$ if, for any
$s,t\in S$, there exists $g\in\Gamma$, such that $g\left(  s\right)  =t$.

\begin{theorem}
\label{petersen}Let $G$ be a Petersen graph. Then $\sigma(G)$ is either
separable or entangled in $\mathbb{C}^{2}\otimes\mathbb{C}^{5}$, depending on
the labelling of $G$.
\end{theorem}

\begin{proof}
Let $G$ (left) and $H$ (right) be the graphs in Figure 4:
\begin{center}
\includegraphics[
height=1.4408in,
width=3.5206in
]%
{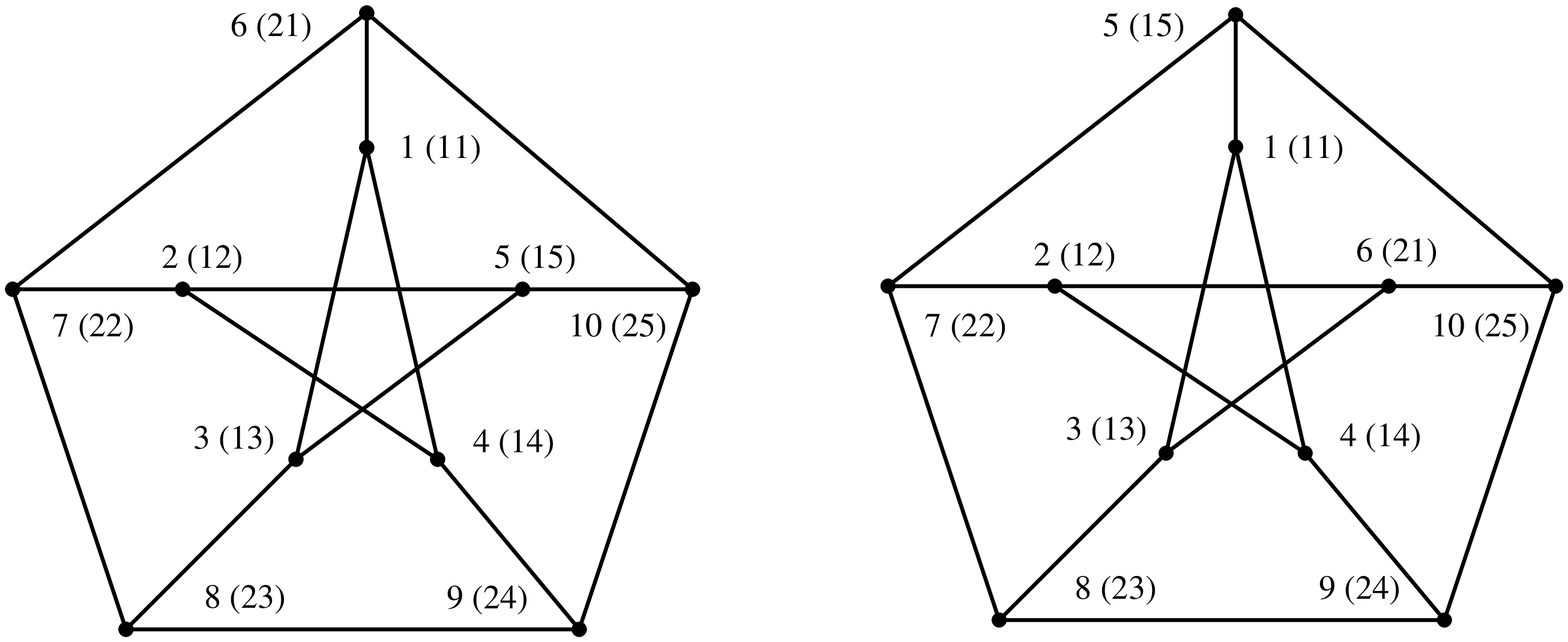}%
\\
Figure 4
\end{center}
Both, $G$ and $H$ are isomorphic to the Petersen graph. The density matrix
$\sigma(G)$ is separable, since every edge of $G$ is separable. The density
matrix of $H$ is entangled in $\mathbb{C}_{A}^{2}\otimes\mathbb{C}_{B}^{5}$,
since it can be shown that $(\sigma(H))^{\intercal_{B}}\ngeqq0$.
\end{proof}

\begin{corollary}
\label{copetersen}The density matrices of strongly-regular graphs and
transitive graphs can be separable or entangled.
\end{corollary}

\begin{proof}
By Theorem \ref{petersen}, since the Petersen graph is strongly-regular and transitive.
\end{proof}

\subsection{Concurrence}

Let $|\psi\rangle_{AB}\in\mathbb{C}_{A}^{2}\otimes\mathbb{C}_{B}^{2}$. The
notion of concurrence was introduced by Wootters \cite{W98}. The
\emph{concurrence} of $|\psi\rangle_{AB}$ is denoted and defined as follows:
$\mathcal{C(}\psi)=\sqrt{2(1-\text{tr}(\rho_{A}^{2}))}$, where $\rho_{A}%
=$tr$_{B}(|\psi\rangle_{AB}\langle\psi|)$. Let $\rho_{AB}$ be a density matrix
acting on $\mathbb{C}_{A}^{2}\otimes\mathbb{C}_{B}^{2}$. The concurrence of
$\rho_{AB}$ is denoted and defined as follows: $\mathcal{C(}\rho_{AB}%
)=\inf\{\sum_{i}\omega_{i}\mathcal{C(}\psi_{i}):\rho_{AB}=\sum_{i}\omega
_{i}|\psi_{i}\rangle_{AB}\langle\psi_{i}|$, where $0\leq\omega_{i}\leq
1,\sum_{i}\omega_{i}=1\}$. Let $\sigma_{y}=-i|1\rangle\langle2|+i|2\rangle
\langle1|$, where $|1\rangle$ and $|2\rangle$ are the eigenvectors of the
matrix
\[
\sigma_{z}=\left[
\begin{array}
[c]{cc}%
1 & 0\\
0 & -1
\end{array}
\right]  ,
\]
corresponding to the eigenvalues $1$ and $-1$, respectively. Let $M^{\ast}$ be
the conjugate of a complex matrix $M$. An analytical formula for
$\mathcal{C(}\rho_{AB})$, is given by $\mathcal{C(}\rho_{AB})=\max
\{0,\lambda_{1}-\lambda_{2}-\lambda_{3}-\lambda_{4}\}$, where $\lambda
_{1},\lambda_{2},\lambda_{3}$ and $\lambda_{4}$ are the square roots of the
eigenvalues of $\rho_{AB}\widetilde{\rho}_{AB}$ arranged in decreasing order,
and $\widetilde{\rho}_{AB}:=(\sigma_{y}\otimes\sigma_{y})\rho_{AB}^{\ast
}(\sigma_{y}\otimes\sigma_{y})$. There are $12$ nonisomorphic graphs on $4$
vertices. Seven of these graphs have entangled density matrix, independently
of the labeling. In the table below, the graphs and the respective concurrence
are given. Note that in three cases the value of the concurrence is exactly fractional.%

\[%
\begin{tabular}
[c]{|c|c|c|c|c|c|c|c|}\hline
Concurrence & $0.33326668$ & $1/3$ & $1/5$ & $1$ & $0.25005352$ &
$0.500131893$ & $0.333236542$\\\hline
Graph &
{\includegraphics[
height=0.4465in,
width=0.3609in
]%
{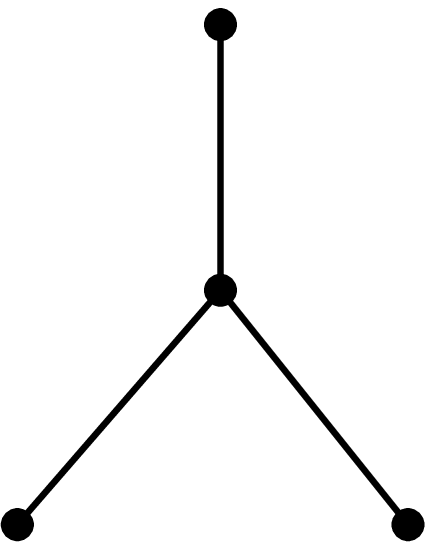}%
}%
&
{\includegraphics[
height=0.2626in,
width=0.3651in
]%
{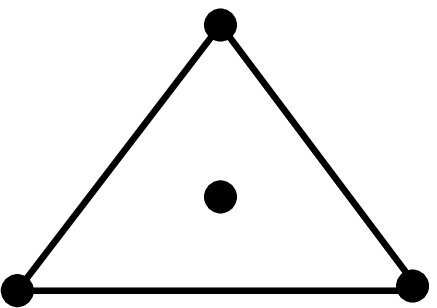}%
}%
&
{\includegraphics[
height=0.5202in,
width=0.3728in
]%
{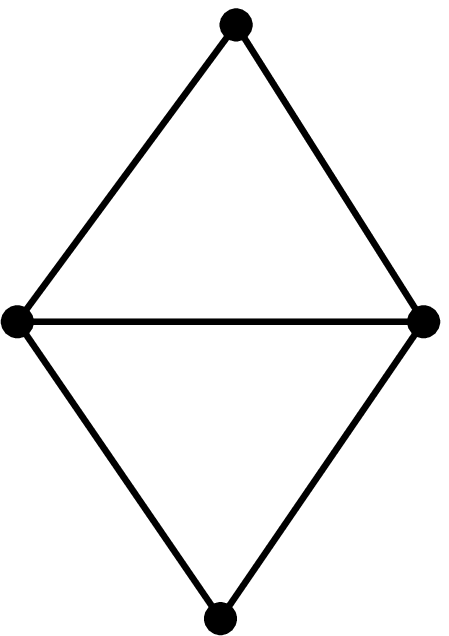}%
}%
&
{\includegraphics[
height=0.227in,
width=0.3516in
]%
{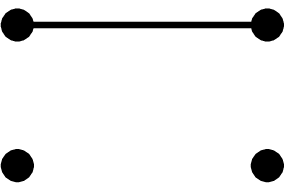}%
}%
&
{\includegraphics[
height=0.46in,
width=0.3228in
]%
{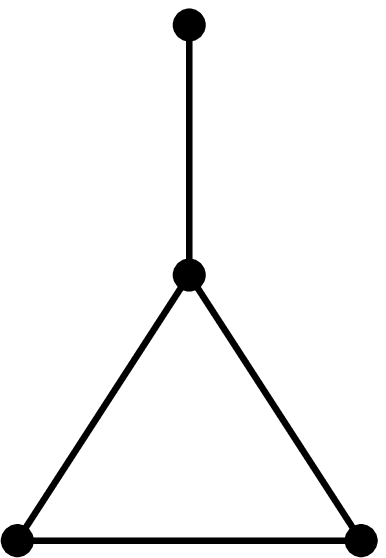}%
}%
&
{\includegraphics[
height=0.2728in,
width=0.4109in
]%
{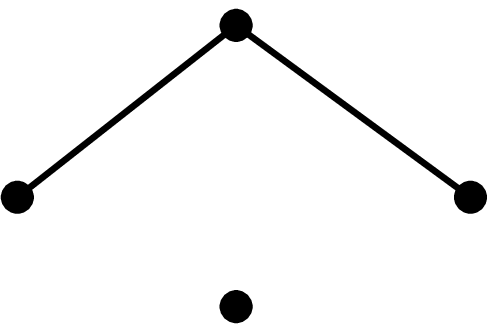}%
}%
&
{\includegraphics[
height=0.2236in,
width=0.4592in
]%
{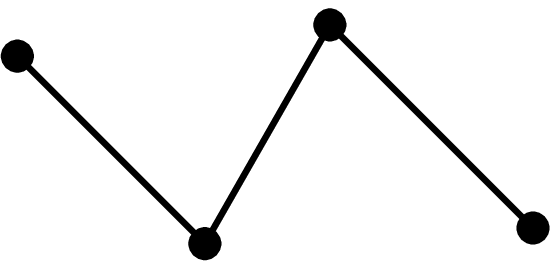}%
}%
\\\hline
\end{tabular}
\]

\section{Graph operations}

A \emph{graph operation} is a map that takes a graph to another one. In Graph
Theory, the study of graph operations consists of a vast literature
\cite{Pr95}. The following are two examples of graph operations.

\begin{example}
\emph{Deleting an edge }$\{v_{i},v_{j}\}$ from a graph $G$ means to transform
$G$ into the graph $G-\{v_{i},v_{j}\}\overset{def}{=}(V(G),E(G)\backslash
\{v_{i},v_{j}\})$. \emph{Adding an edge }$\{v_{i},v_{j}\}$ to a graph $G$,
where $\{v_{i},v_{j}\}\notin E(G)$, means to transform $G$ into the graph
$G+\{v_{i},v_{j}\}\overset{def}{=}(V(G),E(G)\cup\{v_{i},v_{j}\})$.
\emph{Deleting a vertex }$v_{i}$ from a graph $G$ means to transform $G$ into
the graph $G-v_{i}\overset{def}{=}(V(G)\backslash\{v_{i}\},E(G)\backslash
E_{i})$, where $E_{i}$ is the set of all edges incident to $v_{i}$.
\emph{Adding a vertex }$v_{i}$ to a graph $G$ means to transform $G$ into the
graph $G+v_{i}+T_{i}\overset{def}{=}(V(G)\cup\{v_{i}\},E(G)\cup T_{i})$, where
$T_{i}$ is an arbitrary set of edges incident to $v_{i}$.
\end{example}

Let $\mathcal{B}\left(  \mathcal{H}^{n}\right)  $ be the space of all bounded
linear operators on $\mathcal{H}^{n}$. A linear map $\Lambda:\mathcal{B}%
\left(  \mathcal{H}^{n}\right)  \longrightarrow\mathcal{B}\left(
\mathcal{H}^{m}\right)  $ is said to be \emph{hermiticity preserving} if for
every hermitian operator $O\in\mathcal{B}\left(  \mathcal{H}^{n}\right)  $,
$\Lambda(O)$ is an hermitian operator in $\mathcal{B}\left(  \mathcal{H}%
^{m}\right)  $. A hermiticity preserving map $\Lambda:\mathcal{B}\left(
\mathcal{H}^{n}\right)  \longrightarrow\mathcal{B}\left(  \mathcal{H}%
^{m}\right)  $ is said to be \emph{positive} if for any positive operator
$O\in\mathcal{B}\left(  \mathcal{H}^{n}\right)  $, $\Lambda\left(  O\right)  $
is a positive operator in $\mathcal{B}\left(  \mathcal{H}^{m}\right)  $. A
positive map $\Lambda:\mathcal{B}\left(  \mathcal{H}^{n}\right)
\longrightarrow\mathcal{B}\left(  \mathcal{H}^{m}\right)  $ is said to be
\emph{completely positive} if for each positive integer $k$, $(\Lambda\otimes
I_{k^{2}}):\mathcal{B(H}^{n}\otimes\mathcal{H}^{k})\longrightarrow
\mathcal{B(H}^{m}\otimes\mathcal{H}^{k})$ is again a positive map. A
completely positive map $\Lambda:\mathcal{B}\left(  \mathcal{H}^{n}\right)
\longrightarrow\mathcal{B}\left(  \mathcal{H}^{m}\right)  $ is said to be
\emph{trace preserving} if tr$\left(  \Lambda\left(  O\right)  \right)  =$
tr$\left(  O\right)  $, for all $O\in\mathcal{B}\left(  \mathcal{H}%
^{n}\right)  $. A \emph{quantum operation} is a trace preserving completely
positive map (for short, TPCP). In Standard Quantum Mechanics, any physical
transformation of a quantum mechanical system is described by a quantum
operation. We are going to use the following result:

\noindent\textbf{(Kraus representation theorem \label{Kxx})} Given a quantum
operation $\Lambda:\mathcal{B}\left(  \mathcal{H}^{n}\right)  \longrightarrow
\mathcal{B}\left(  \mathcal{H}^{m}\right)  $, there exist $n\times m$ matrices
$A_{i}$, such that $\Lambda\left(  \rho\right)  =\sum_{i}A_{i}\rho
A_{i}^{\dagger}$, where $\rho$ is any density matrix acting on $\mathcal{H}%
^{n}$ and $\sum_{i}A_{i}^{\dagger}A_{i}=I_{m}$. (The converse is also true.)
The matrices $A_{i}$'s are called \emph{Kraus operators}.

A \emph{projective measurement }$\mathcal{M}=\{P_{i}:i=1,2,...,n\}$, on a
quantum mechanical system $\emph{S}$ whose state is $\rho$, consists of
pairwise orthogonal projectors $P_{i}:\mathcal{H}_{\emph{S}}\longrightarrow
\mathcal{H}_{\emph{S}}$, such that $\sum_{i=1}^{n}P_{i}=I_{\dim(\mathcal{H}%
_{\emph{S}})}$. The $i$-th outcome of the measurement occurs with probability
tr$(P_{i}\rho)$ and the post-measurement state of $\emph{S}$ is $\frac
{P_{i}\rho P_{i}}{\text{tr}(P_{i}\rho)}$. Whenever the $i$-th outcome of the
measurement occurs, we say that $P_{i}$ \emph{clicks}.

\subsection{Deleting and adding an edge\label{delee}}

Here we describe how to delete or add an edge by means of TPCP. Let $G$ be a
graph on $n$ vertices, $v_{1},v_{2},...,v_{n}$, and $m$ edges, $\{v_{i_{1}%
},v_{j_{1}}\},\{v_{i_{2}},v_{j_{2}}\},...,\{v_{i_{m}},v_{j_{m}}\}$, where
$1\leq i_{1},j_{1},i_{2},j_{2},...,i_{m},j_{m}\leq n$. Our purpose is to
delete the edge $\{v_{i_{k}},v_{j_{k}}\}$. Then we have%
\[
\sigma(G)=\frac{1}{m}\sum_{l=1}^{m}\sigma(H_{i_{l}j_{l}})=\frac{1}{m}%
\sum_{l=1}^{m}P[\frac{1}{\sqrt{2}}\left(  \left\vert v_{i_{l}}\right\rangle
-\left\vert v_{j_{l}}\right\rangle \right)  ]
\]
and%
\[
\sigma(G-\{v_{i_{k}},v_{j_{k}}\})=\frac{1}{m-1}\sum_{l=1,l\neq k}^{m}%
\sigma(H_{i_{l}j_{l}})=\frac{1}{m-1}\sum_{l=1,l\neq k}^{m}P[\frac{1}{\sqrt{2}%
}\left(  \left\vert v_{i_{l}}\right\rangle -\left\vert v_{j_{l}}\right\rangle
\right)  ].
\]
A measurement in the following basis, $\mathcal{M}=\{\frac{1}{\sqrt{2}}\left(
\left\vert v_{i_{k}}\right\rangle \pm\left\vert v_{j_{k}}\right\rangle
\right)  ,|v_{i}\rangle:i\neq i_{k},j_{k}$ and $i=1,2,...,n\}$ is performed on
the system prepared in the state $\sigma(G)$. The probability that $P[\frac
{1}{\sqrt{2}}\left(  \left\vert v_{i_{k}}\right\rangle +\left\vert v_{j_{k}%
}\right\rangle \right)  ]$ clicks is
\begin{equation}
\frac{m-1}{2m}\left(  \langle v_{i_{k}}|+\langle v_{j_{k}}|\right)
\sigma(G-\{v_{i_{k}},v_{j_{k}}\})(\left\vert v_{i_{k}}\right\rangle
+\left\vert v_{j_{k}}\right\rangle )=\frac{1}{4m}\sum_{l=1,l\neq k}^{m}%
(\delta_{i_{k}i_{l}}-\delta_{i_{k}j_{l}}+\delta_{j_{k}i_{l}}-\delta
_{j_{k}j_{l}})^{2}. \label{prob}%
\end{equation}
The state after the measurement is $P[\frac{1}{\sqrt{2}}\left(  \left\vert
v_{i_{l}}\right\rangle +\left\vert v_{j_{l}}\right\rangle \right)  ]$. Let
$U_{kl}^{+}$ be an $n\times n$ unitary matrix, such that $U_{kl}^{+}\frac
{1}{\sqrt{2}}\left(  \left\vert v_{i_{k}}\right\rangle +\left\vert v_{j_{k}%
}\right\rangle \right)  =\frac{1}{\sqrt{2}}\left(  \left\vert v_{i_{l}%
}\right\rangle -\left\vert v_{j_{l}}\right\rangle \right)  $, for
$l=1,2,...,k-1,k+1,...,m$. Now, with probability $1/\left(  m-1\right)  $ we
apply $U_{kl}^{+}$ on $P[\frac{1}{\sqrt{2}}\left(  \left\vert v_{i_{k}%
}\right\rangle +\left\vert v_{j_{k}}\right\rangle \right)  ]$, for each
$l=1,2,...,k-1,k+1,...,m$. Finally we obtain $\sigma(G-\{v_{i_{k}},v_{j_{k}%
}\})$ with probability given by Equation \ref{prob}. The probability that
$P[\frac{1}{\sqrt{2}}\left(  \left\vert v_{i_{k}}\right\rangle -\left\vert
v_{j_{k}}\right\rangle \right)  ]$ clicks is
\begin{equation}
\frac{1}{4m}\sum_{l=1}^{m}(\delta_{i_{k}i_{l}}-\delta_{i_{k}j_{l}}%
-\delta_{j_{k}i_{l}}+\delta_{j_{k}j_{l}})^{2}. \label{prob1}%
\end{equation}
The state after the measurement is $P[\frac{1}{\sqrt{2}}\left(  \left\vert
v_{i_{l}}\right\rangle -\left\vert v_{j_{l}}\right\rangle \right)  ]$. Let
$U_{kl}^{-}$ be an $n\times n$ unitary matrix, such that $U_{kl}^{-}\frac
{1}{\sqrt{2}}\left(  \left\vert v_{i_{k}}\right\rangle -\left\vert v_{j_{k}%
}\right\rangle \right)  =\frac{1}{\sqrt{2}}\left(  \left\vert v_{i_{l}%
}\right\rangle -\left\vert v_{j_{l}}\right\rangle \right)  $, for
$l=1,2,...,k-1,k+1,...,m$. With probability $1/\left(  m-1\right)  $ we apply
$U_{kl}^{-}$ on $P[\frac{1}{\sqrt{2}}\left(  \left\vert v_{i_{k}}\right\rangle
-\left\vert v_{j_{k}}\right\rangle \right)  ]$, for each
$l=1,2,...,k-1,k+1,...,m$. Finally, we obtain $\sigma(G-\{v_{i_{k}},v_{j_{k}%
}\})$ with probability given by Equation \ref{prob1}. The probability that
$P[|v_{i}\rangle]$, where $i\neq i_{k},j_{k}$ and $i=1,2,...,n$, clicks is
\begin{equation}
\frac{1}{2m}\sum_{l=1}^{m}(\delta_{ii_{l}}-\delta_{ij_{l}})^{2} \label{prob2}%
\end{equation}
and the state after the measurement is $P[|v_{i}\rangle]$. Let $U_{il}$ be an
$n\times n$ unitary matrix, such that $U_{il}|v_{i}\rangle=\frac{1}{\sqrt{2}%
}\left(  \left\vert v_{i_{l}}\right\rangle -\left\vert v_{j_{l}}\right\rangle
\right)  $, where for $l=1,2,...,k-1,k+1,...,m$. With probability $1/\left(
m-1\right)  $ we apply $U_{il}$ on $P[|v_{i}\rangle]$, for each
$l=1,2,...,k-1,k+1,...,m$. We obtain $\sigma(G-\{v_{i_{k}},v_{j_{k}}\})$ with
probability given by Equation \ref{prob2}. This completes the process. The set
of Kraus operators that realize the TPCP for deleting the edge $\{v_{i_{k}%
},v_{j_{k}}\}$ is then%
\[%
\begin{tabular}
[c]{ll}
& $\{\frac{1}{\sqrt{m-1}}U_{kl}^{+}P[\frac{1}{\sqrt{2}}\left(  \left\vert
v_{i_{k}}\right\rangle +\left\vert v_{j_{k}}\right\rangle \right)
]:l=1,2,...,k-1,k+1,...,m\}$\\
$\cup$ & $\{\frac{1}{\sqrt{m-1}}U_{kl}^{-}P[\frac{1}{\sqrt{2}}\left(
\left\vert v_{i_{k}}\right\rangle -\left\vert v_{j_{k}}\right\rangle \right)
]:l=1,2,...,k-1,k+1,...,m\}$\\
$\cup$ & $\{\frac{1}{\sqrt{m-1}}U_{il}P[|v_{i}\rangle]:i=1,2,...,n;i\neq
i_{k},j_{k};l=1,2,...,k-1,k+1,...,m\}.$%
\end{tabular}
\ \
\]

The set of Kraus operators that realize the TPCP for adding back the edge
$\{v_{i_{k}},v_{j_{k}}\}$ to $G-\{v_{i_{k}},v_{j_{k}}\}$ is%
\[%
\begin{tabular}
[c]{ll}
& $\{\frac{1}{\sqrt{m}}V_{kl}^{+}P[\frac{1}{\sqrt{2}}\left(  \left\vert
v_{i_{k}}\right\rangle +\left\vert v_{j_{k}}\right\rangle \right)
]:l=1,2,...,m\}$\\
$\cup$ & $\{\frac{1}{\sqrt{m}}V_{kl}^{-}P[\frac{1}{\sqrt{2}}\left(  \left\vert
v_{i_{k}}\right\rangle -\left\vert v_{j_{k}}\right\rangle \right)
]:l=1,2,...,m\}$\\
$\cup$ & $\{\frac{1}{\sqrt{m}}V_{il}P[|v_{i}\rangle]:i=1,2,...,n;i\neq
i_{k},j_{k};l=1,2,...,m\},$%
\end{tabular}
\]
where $V_{kl}^{+}$, $V_{kl}^{-}$ and $V_{il}$ are $n\times n$ unitary matrices
defined as follows:
\[%
\begin{tabular}
[c]{lll}%
$V_{kl}^{+}\frac{1}{\sqrt{2}}\left(  \left\vert v_{i_{k}}\right\rangle
+\left\vert v_{j_{k}}\right\rangle \right)  =\frac{1}{\sqrt{2}}\left(
\left\vert v_{i_{l}}\right\rangle -\left\vert v_{j_{l}}\right\rangle \right)
,$ &  & for $l=1,2,...,m;$\\
$V_{kl}^{-}\frac{1}{\sqrt{2}}\left(  \left\vert v_{i_{k}}\right\rangle
-\left\vert v_{j_{k}}\right\rangle \right)  =\frac{1}{\sqrt{2}}\left(
\left\vert v_{i_{l}}\right\rangle -\left\vert v_{j_{l}}\right\rangle \right)
,$ &  & for $l=1,2,...,m;$\\
$V_{il}\left\vert v_{i}\right\rangle =\frac{1}{\sqrt{2}}\left(  \left\vert
v_{i_{l}}\right\rangle -\left\vert v_{j_{l}}\right\rangle \right)  ,$ &  & for
$i=1,2,...,n;i\neq i_{k},j_{k};l=1,2,...,m.$%
\end{tabular}
\]

\subsection{Deleting and adding a vertex}

Here we describe how to delete or add a vertex by means of TPCP. Let $G$ be a
graph on $n$ vertices, $v_{1},v_{2},...,v_{n}$, and $m$ edges, $\{v_{i_{1}%
},v_{j_{1}}\},\{v_{i_{2}},v_{j_{2}}\},...,\{v_{i_{m}},v_{j_{m}}\}$, where
$1\leq i_{1},j_{1},i_{2},j_{2},...,i_{m},j_{m}\leq n$. Our purpose is to
delete a vertex $v_{i}$. We first delete all the edges incident to $v_{i}$
(cfr. Section \ref{delee}). In this way, we obtain a new graph, say $H$. We
then perform the following projective measurement on $\sigma(H)$:
$\mathcal{M}=\{I_{n}-P[|v_{i}\rangle],P[|v_{i}\rangle]\}$. Given that,
possible loops in $H$ do not appear on $\sigma(H)$, when $\mathcal{M}$ is
performed on $\sigma(H)$, $I_{n}-P[|v_{i}\rangle$ clicks with probability one.
The state after the measurement is $\sigma(G-v_{i})$, which is the state of
the desired graph. Let $G$ be a graph on $n$ vertices, $v_{1},v_{2},...,v_{n}%
$, and $m$ edges, $\{v_{i_{1}},v_{j_{1}}\},\{v_{i_{2}},v_{j_{2}}%
\},...,\{v_{i_{m}},v_{j_{m}}\}$, where $1\leq i_{1},j_{1},i_{2},j_{2}%
,...,i_{m},j_{m}\leq n$. Our purpose is to obtain the graph $G+v_{i}%
=G\uplus\{x\}$. Consider the following density matrix $\rho=(\frac{1}{2}%
\sum_{i=1,2}P[|u_{i}\rangle])\otimes\sigma(G)$, where $\{|u_{1}\rangle
,|u_{2}\rangle\}$ forms an orthonormal basis of $\mathbb{C}^{2}$. We associate
the vertex $u_{i}$ to the state $|u_{i}\rangle$ for $i=1,2$. Consider the
graph $H=(\{u_{1},u_{2}\},\{\{u_{1},u_{1}\},\{u_{2},u_{2}\}\})$. It is easy to
check (cfr. Equation \ref{newlap}) that $\sigma_{\circ}(H)=\frac{1}{2}%
\sum_{i=1,2}P[|u_{i}\rangle]$. Also observe that $\rho=\sigma(H\otimes G)$.
Thus $H\otimes G$ is a graph on $2n$ vertices labeled by $u_{1}v_{1}%
,u_{1}v_{2},...,u_{1}v_{n},u_{2}v_{1},u_{2}v_{2},...,u_{2}v_{n}$ and with $2m$
edges
\[
\{u_{1}v_{i_{1}},u_{1}v_{j_{i}}\},...,\{u_{1}v_{i_{m}},u_{1}v_{j_{m}%
}\},\{u_{2}v_{i_{1}},u_{2}v_{j_{i}}\},...,\{u_{2}v_{i_{m}},u_{2}v_{j_{m}}\}.
\]
So, $H\otimes G=H_{1}\uplus H_{2}$, where $H_{1}=(\{u_{1}v_{1},...,u_{1}%
v_{n}\},\{\{u_{1}v_{i_{1}},u_{1}v_{j_{i}}\},...,\{u_{1}v_{i_{m}},u_{1}%
v_{j_{m}}\}\})$ and $H_{2}=\{\{u_{2}v_{1},...,u_{2}v_{n}\},\{\{u_{2}v_{i_{1}%
},u_{2}v_{j_{i}}\},...,\{u_{2}v_{i_{m}},u_{2}v_{j_{m}}\}\}\}$. We first delete
all the edges of $H\otimes G$ which are incident to the vertex $u_{2}v_{1}\in
V(H_{2})$. Now, we perform the following projective measurement on
$\sigma(G\otimes H)$:%
\[
\mathcal{M}=\{I_{2n}-\sum_{i=2}^{n}P[|u_{2}\rangle|v_{i}\rangle],\sum
_{i=2}^{n}P[|u_{2}\rangle|v_{i}\rangle]\}.
\]
The probability that $I_{2n}-\sum_{i=2}^{n}P[|u_{2}\rangle|v_{i}\rangle]$
clicks is one and the state after the measurement is $\sigma(H_{1}+u_{2}%
v_{1})$, where $H_{1}\cong G$.

\subsection{LOCC}

A \emph{local operation and classical communication} (for short, \emph{LOCC})
is a TPCP $\Lambda:\mathcal{B}\left(  \mathcal{H}_{A}\otimes\mathcal{H}%
_{B}\right)  \longrightarrow\mathcal{B}\left(  \mathcal{K}_{A}\otimes
\mathcal{K}_{B}\right)  $ defined in the following way. The TPCP $\Lambda$ is
an LOCC if, for some $n>0$, there exist sequences of Hilbert spaces
$(\mathcal{H}_{A}^{k})_{k=1}^{n+1}$ and $(\mathcal{H}_{B}^{k})_{k=1}^{n+1}$
with $\mathcal{H}_{A}^{1}=\mathcal{H}_{A}$, $\mathcal{H}_{B}^{1}%
=\mathcal{H}_{B}$, $\mathcal{H}_{A}^{n+1}=\mathcal{K}_{A}$ and $\mathcal{H}%
_{B}^{n+1}=\mathcal{K}_{B}$, such that $\Lambda$ can be written in the
following form $\Lambda(\sigma)=\sum_{i_{1},...,i_{2n}=1}^{K_{1},....,K_{2n}%
}V_{i_{1},...,i_{2n}}^{AB}\sigma(V_{i_{1},...,i_{2n}}^{AB})^{\dagger}$ for all
$\sigma\in\mathcal{B}\left(  \mathcal{H}_{A}\otimes\mathcal{H}_{B}\right)  $.
Let $I_{A}^{n}:\mathcal{H}_{A}^{n}\longrightarrow\mathcal{H}_{A}^{n}$ and
$I_{B}^{n}:\mathcal{H}_{B}^{n}\longrightarrow\mathcal{H}_{B}^{n}$ be identity
operators. In $\Lambda(\sigma)$, $V_{i_{1},...,i_{2n}}^{AB}:\mathcal{H}%
_{A}\otimes\mathcal{H}_{B}\longrightarrow\mathcal{K}_{A}\otimes\mathcal{K}%
_{B}$ is given by%
\[
V_{i_{1},...,i_{2n}}^{AB}\overset{def}{=}(I_{A}^{n+1}\otimes W_{2n}%
^{i_{2n},...,i_{1}})(V_{2n-1}^{i_{2n-1,...,i_{1}}}\otimes I_{B}^{n})(I_{A}%
^{n}\otimes W_{2n-2}^{i_{2n-2},...,i_{1}})....(I_{A}^{2}\otimes W_{2}%
^{i_{2},i_{1}})(V_{1}^{i_{1}}\otimes I_{B}^{1}).
\]
The sequences $(V_{2k-1}^{i_{2k-1,...,i_{1}}}:\mathcal{H}_{A}^{k}%
\longrightarrow\mathcal{H}_{A}^{k+1})_{k=1}^{n}$ and $(W_{2k}^{i_{2k,...,i_{1}%
}}:\mathcal{H}_{B}^{k}\longrightarrow\mathcal{H}_{B}^{k+1})_{k=1}^{n}$ are
families of operator. For each sequence of indices $(i_{2k},...,i_{1})$ and
for $k=0,1,...,n-1$, $\sum_{i_{2k+1}=1}^{K_{2n+1}}(V_{2k+1}^{i_{2k+1,...,i_{1}%
}})^{\dagger}V_{2k+1}^{i_{2k+1,...,i_{1}}}=I_{A}^{k+1}$, for each sequence of
indices $(i_{2k-1},...,i_{1})$ and for $k=0,1,...,n$, $\sum_{i_{2k+2}%
=1}^{K_{2k+2}}(W_{2k+2}^{i_{2k,...,i_{1}}})^{\dagger}W_{2k+2}^{i_{2k,...,i_{1}%
}}=I_{B}^{k}$.

\noindent\textbf{Thermodynamic Principle. }One can not obtain an entangled
state from a separable state by using LOCC.

A consequence of this principle is that, given two (possibly isomorphic)
graphs $G$ and $H$ on $n=pq$ vertices, we can always obtain $\sigma(H)$ from
$\sigma(G)$ by using LOCC only, if $\sigma(G)$ is separable or entangled and
$\sigma(H)$ is separable, in $\mathbb{C}^{p}\otimes\mathbb{C}^{q}$.

\begin{example}
\label{loccex}Let $G\cong2K_{2}$ and let $\{11,22\},\{12,21\}\in E(G)$. Then%
\[
\sigma(G)=\frac{1}{2}P[\frac{1}{\sqrt{2}}[|1\rangle|1\rangle-[|2\rangle
|2\rangle]+\frac{1}{2}P[\frac{1}{\sqrt{2}}[|1\rangle|2\rangle-[|2\rangle
|1\rangle].
\]
This density matrix is separable. Can we delete an edge of $G$ by LOCC? The
answer is no. If we can delete $\{12,21\}$ (or, equivalently, $\{11,22\}$) by
LOCC, we obtain $\sigma(G-\{12,21\})=P[\frac{1}{\sqrt{2}}[|1\rangle
|1\rangle-[|2\rangle|2\rangle]$, which is entangled. This fact violates the
thermodynamic principle.
\end{example}

\begin{example}
Let $G\cong K_{4}-e$, for some edge $e$. Let $f$ be the edge of $G$ incident
with the vertices of degree $3$. Then $\sigma(G-f)$ is separable independent
of the labeling. From $G$ we can always obtain $G-f$ by LOCC.
\end{example}

\begin{example}
Lemma \ref{complete} together with Theorem \ref{star} and the thermodynamic
principle, show that we can not obtain $K_{1,n-1}$ from $K_{n}$ by LOCC.
\end{example}

\section{Open problems}

\begin{problem}
The separability of $K_{1,n-1}$ and $K_{n}$ do not depend on their labeling.
Are these the only classes of graphs for which this happens? In general, give
\emph{separability criteria} for density matrices of graphs.
\end{problem}

\begin{problem}
Let $\sigma(G)$ be entangled in $\mathbb{C}^{p}\otimes\mathbb{C}^{q}$. In
general, whether a graph operation on $G$ can be implemented by an LOCC
depends on $G$ and on its labeling. The following are natural questions: (1)
What are the most general conditions on $G$ and on its labeling such that a
graph $H$ can be obtained from $G$ by LOCCs? (2) Does there exist a graph
operation implemented by an LOCC independent of the labeling? (3) Given a
graph $G$, with specific properties, determine the set of all graphs which are
obtainable from $G$ by means of LOCCs.
\end{problem}

\begin{problem}
Studying the realization of TPCP in relation to the tensor product of graphs.
\end{problem}

\begin{problem}
We have calculated the concurrence of density matrices of graphs entangled in
$\mathbb{C}^{2}\otimes\mathbb{C}^{2}$. It turns out that for some graphs $G$
the concurrence is equal to $\frac{1}{|E|}$. For some other graphs the
concurrence is $\frac{1}{|E|}\pm\varepsilon$. Are these observations related
to some property of the graphs?
\end{problem}

\begin{conjecture}
Let $G$ be a graph ($|V|=pq$). If $G$ has only one entangled edge then
$\sigma(G)$ is entangled; if all the entangled edges of $G$ are incident to
the same vertex then $\sigma(G)$ is entangled.
\end{conjecture}

Let $\rho_{AB}$ be a density matrix acting on $\mathbb{C}_{A}^{p}%
\otimes\mathbb{C}_{B}^{q}$, where $pq=n$. Let $S_{\rho}=\{\{p_{i},|\psi
_{i}\rangle:i=1,2,...,N\}:\rho_{AB}=\sum_{i=1}^{n}p_{i}|\psi_{i}\rangle
_{AB}\langle\psi_{i}|$, where $|\psi_{i}\rangle_{AB}\in\mathbb{C}_{A}%
^{p}\otimes\mathbb{C}_{B}^{q},0\leq p_{i}\leq1$ and $\sum_{i=1}^{N}p_{i}=1\}.$
The \emph{entanglement of formation} of $\rho_{AB}$ is denoted and defined by
$E_{F}(\rho_{AB})=\inf_{\{p_{i},|\psi_{i}\rangle:i=1,2,...,N\}\in S_{\rho}%
}\sum_{i=1}^{N}p_{i}S(tr_{X}(|\psi_{i}\rangle_{AB}\langle\psi_{i}|))$, where
$X=A$ or $X=B$.

\begin{conjecture}
Let $G$ be a graph ($|V|=pq$) with $m$ edges. If $\sigma(G)$ is entangled in
$\mathbb{C}_{A}^{p}\otimes\mathbb{C}_{B}^{q}$ then $E_{F}(\sigma
(G))\approx\frac{1}{m}\sum_{k=1}^{m}E_{F}(\sigma(k))$, where $\sigma(k)$ is
the pure density matrix associated to the $k$-th edge of $G$.
\end{conjecture}

\begin{conjecture}
Let $\mathcal{G}_{n}^{c}$ be the set of all connected graphs on $n$ vertices.
Let $G\in\mathcal{G}_{n}^{c}$ ($|V|=pq$). Then $\max_{\mathcal{G}_{n}^{c}%
}E_{F}(\sigma(G))=E_{F}(\sigma(K_{1,n-1}))$.
\end{conjecture}


\begin{thebibliography}{99}                                                                                               %


\bibitem {BCPP02}J. Batle, M. Casas, A. R. Plastino and A. Plastino,
Entanglement, mixedness, and $q$-entropies, \emph{Phys. Lett. A} \textbf{296}
(2002), \emph{no. 6}, 251--258.

\bibitem {GR01}C. Godsil and G. Royle, Algebraic graph theory. Graduate Texts
in Mathematics, 207. Springer-Verlag, New York, 2001.

\bibitem {HHH96}M. Horodecki, P. Horodecki, R. Horodecki, Separability of
mixed states: necessary and sufficient conditions, \emph{Phys. Lett. A}
\textbf{223} (1996), \emph{no. 1-2}, 1--8.

\bibitem {HHH98}M. Horodecki, P. Horodecki, R. Horodecki, Mixed-state
entanglement and distillation: Is there a "bound" entanglement in nature?
\emph{Phys. Rev. Lett.} \textbf{80} (1998), \emph{no. 24}, 5239--5242.

\bibitem {IK00}W. Imrich and S. Klav\v{z}ar, \emph{Product graphs. Structure
and recognition}. With a foreword by Peter Winkler. Wiley-Interscience Series
in Discrete Mathematics and Optimization. Wiley-Interscience, New York, 2000.

\bibitem {KMW03}J. P. Keating, J. Marklof and B. Winn, Value distribution of
the eigenfunctions and spectral determinants of quantum star graphs.
\emph{Comm. Math. Phys.} \textbf{241} (2003), \emph{no. 2-3}, 421--452.

\bibitem {M91}B. Mohar, \emph{The Laplacian spectrum of graphs}. Graph theory,
combinatorics, and applications. Vol. 2 (Kalamazoo, MI, 1988), 871--898,
Wiley-Intersci. Publ., Wiley, New York, 1991.

\bibitem {P93}A Peres, \emph{Quantum theory: concepts and methods}.
Fundamental Theories of Physics, 57. Kluwer Academic Publishers Group,
Dordrecht, 1993.

\bibitem {P96}A. Peres, Separability criterion for density matrices.
\emph{Phys. Rev. Lett.} \textbf{77} (1996), \emph{no. 8}, 1413--1415.

\bibitem {Pr95}E. Prisner, \emph{Graph dynamics}. Pitman Research Notes in
Mathematics Series, 338. Longman, Harlow, 1995.

\bibitem {SHK02}J.-L. Shu, Y. Hong and W.-R. Kai, A sharp upper bound on the
largest eigenvalue of the Laplacian matrix of a graph. \emph{Linear Algebra
Appl.} \textbf{347} (2002), 123--129.

\bibitem {W98}W. K. Wootters, Entanglement of formation and concurrence.
\emph{Quantum Inf. Comput.} \textbf{1} (2001), \emph{no. 1}, 27--44.
\end{thebibliography}
\end{document}